\documentclass[aps,prd,preprint,superscriptaddress,nofootinbib,preprintnumbers]{revtex4-1}
\pdfoutput=1 % if your are submitting a pdflatex (i.e. if you have
\usepackage{amsmath,amssymb,amsfonts}
\usepackage{graphicx}
\usepackage{booktabs,colortbl}
\usepackage{cancel,xcolor}
\usepackage[normal]{caption}
\usepackage{subcaption}
\newcommand{\tev}{~\text{TeV}}
\newcommand{\gev}{~\text{GeV}}
\newcommand{\nn}{\nonumber \\ }
\newcommand{\bra}{\langle}
\newcommand{\ket}{\rangle}
\newcommand{\Lc}{\mathcal{L}}

\newcommand{\be}{\begin{equation}} 
\newcommand{\ee}{\end{equation}} 
\newcommand{\bea}{\begin{eqnarray}}  
\newcommand{\eea}{\end{eqnarray}}
\newcommand{\bs}{\begin{split}} 
\newcommand{\es}{\end{split}}

\begin{document}

%----------------------------------- TITLE AND AUTHORS -----------------------------------------%

\title{\sc Dirac Triplet Extension of the MSSM }

\author{C. Alvarado\footnote{E-mail: calvara1@nd.edu}}
\affiliation{Department of Physics, University of Notre Dame\\225 Nieuwland Science Hall, Notre Dame, IN 46556, U.S.A.}
\author{A. Delgado\footnote{E-mail: adelgad2@nd.edu}}
\affiliation{Department of Physics, University of Notre Dame\\225 Nieuwland Science Hall, Notre Dame, IN 46556, U.S.A.}
\affiliation{Theory Division, Physics Department, CERN, CH-1211 Geneva 23, Switzerland}
\author{A. Martin\footnote{E-mail: amarti41@nd.edu} }
\affiliation{Department of Physics, University of Notre Dame\\225 Nieuwland Science Hall, Notre Dame, IN 46556, U.S.A.}
\author{and B. Ostdiek\footnote{E-mail: bostdiek@nd.edu} }
\affiliation{Department of Physics, University of Notre Dame\\225 Nieuwland Science Hall, Notre Dame, IN 46556, U.S.A.}

\preprint{CERN-PH-TH-2015-084}
%--------------------------------------------- ABSTRACT ---------------------------------------------%
 
\begin{abstract}
\vspace*{0.5cm}
In this paper we explore extensions of the Minimal Supersymmetric Standard Model involving two $SU(2)_L$ triplet chiral superfields that share a superpotential Dirac mass yet only one of which couples to the Higgs fields. This choice is motivated by recent work using two singlet superfields with the same superpotential requirements. We find that, as in the singlet case, the Higgs mass in the triplet extension can easily be raised to $125\,\gev$ without introducing large fine-tuning. For triplets that carry hypercharge, the regions of least fine tuning are characterized by small contributions to the $\mathcal T$ parameter, and light stop squarks, $m_{\tilde t_1} \sim 300-450\,\gev$; the latter is a result of the $\tan\beta$ dependence of the triplet contribution to the Higgs mass. Despite such light stop masses, these models are viable provided the stop-electroweakino spectrum is sufficiently compressed.
\end{abstract}

\maketitle
%-------------------------------- DOCUMENT: INTRODUCTION ---------------------------------%
\newpage
\section{Introduction}
\label{sec:intro}

The Minimal Supersymmetric Standard Model (MSSM) sets $m_{Z}$ as the upper bound of the tree-level mass of the lightest {\it CP} even scalar in the spectrum. Since this particle is commonly identified with the Standard Model Higgs boson, either large one-loop corrections due to heavy third family squarks or a high degree of stop mixing are necessary to push $m_{h}$ up to the observed value of $\sim 125 \gev$ \cite{Aad:2012tfa, Chatrchyan:2012ufa}. Either of these two requirements on the stops introduces sub-percent fine tuning \cite{Hall:2011aa}. This occurs because both effects radiatively induce large corrections to the soft mass of the Higgs field $m_{H_u}^2$, which must be canceled off in order stabilize the electroweak scale. In this sense, the observation of the Higgs with a 125 GeV mass 
makes the MSSM alarmingly fine-tuned, independent of the fact that we have not yet discovered any supersymmetric particles.

 A variety of techniques have been proposed to avoid such a heavy stop spectrum. The simplest possibilities are to extend the MSSM gauge group or field content, respectively modifying the $D$- and $F$-terms of the Higgs potential \cite{Cvetic:1997ky, Ellwanger:2009dp, Delgado:2010cw}. While the former necessarily alters the quartic terms in a manner dictated by the gauge group, the later relies on raising the quartic coupling of the Higgses via the inclusion of extra superpotential couplings. 
 
A class of well-known models based on this effect is the Next-to-Minimal Supersymmetric Standard Model (NMSSM) which adds a gauge singlet field $S$ to the MSSM. Although capable of rendering the correct Higgs mass, the NMSSM does so by decoupling the scalar part of the singlet superfield. However, the soft mass of the singlet feeds back into $m^2_{H_u}, m^2_{H_d}$ at one loop via the renormalization group equations. Large singlet masses therefore can lead to large corrections to $m_{H_{u,d}}^2$, so the NMSSM solution to the Higgs mass comes at the expense of substantial tine tuning.  

The authors of \cite{Lu:2013cta} extended the NMSSM with a second singlet $\bar{S}$ which does not couple to the Higgs doublets yet has a superpotential mass term with $S$:
\begin{equation}
W=W_{\text{Yukawa}}+(\mu+\lambda S)H_{u}H_{d}+MS\bar{S},
\end{equation}
where $W_{\text{Yukawa}}$ stands for the usual MSSM Yukawa terms; due to the Dirac mass term between the singlets, the model was dubbed the DiracNMSSM. The tree level Higgs mass squared in this setup is modified, receiving a positive contribution that depends on the $\bar S$ soft mass, and a negative contribution that depends on the $S$ soft mass. Including the one-loop correction from stop loops (see for example Ref~\cite{Carena:2011aa}), the resulting Higgs mass is
\begin{align}
m^2_h = &m_Z^2 \cos^2(2\beta) +  (\text{stop loops}) \nonumber \\
& + \lambda^2 v^2 \sin^2(2\beta)\left(\frac{m_{\bar{S}}^2}{M^2+m_{\bar{S}}^2}\right) - \frac{\lambda^2v^2}{M^2+m_S^2}\left|A_{\lambda} \sin(2\beta) -2\mu^* \right|^2.
\label{eqn:singletModel}
\end{align}

To efficiently raise the Higgs mass, one takes advantage of the positive term while trying to keep the negative term as small as possible. The positive term is increased by taking the soft mass of the non coupled singlet -- $m^2_{\bar S}$ -- to be much larger than the supersymmetric mass term, $M$. If $M$ is also larger than $\lambda^2v^2$ then the negative term is minimized. While large singlet masses in the NMSSM come hand-in-hand with increased tuning, this does not happen here. Specifically, the authors of \cite{Lu:2013cta} showed that the mass of $\bar{S}$ can be raised almost indefinitely without introducing fine tuning -- a clear violation of the conventional wisdom that increases to the Higgs mass require new light states. As explained in \cite{Lu:2013cta}, the keys to this behavior are the Dirac mass term between $S$ and $\bar S$ and the absence of couplings of $\bar{S}$ with $H_{u}, H_{d}$. A detailed study of the DiracNMSSM was performed in Ref.~\cite{Kaminska:2014wia}, taking into account all corrections at one loop order and dominant two-loop corrections. Beyond fine tuning, constraints from SUSY searches and dark matter were also applied.
 
One disadvantage of the original DiracNMSSM is that the singlet contribution to $m^2_h$ has the same $\tan\beta$ dependence as in the NMSSM. Specifically, the singlet piece is largest at low $\tan\beta$, exactly the region where the MSSM tree level Higgs mass vanishes. This can be overcome, but requires sizable coupling of the singlet to Higgses.  

In this paper, we examine the effects of replacing the singlets in the DiracNMSSM with triplets under $SU(2)_L$,  maintaining the key features of the Dirac mass and with only one triplet coupled to the Higgses. Triplet extensions of the MSSM have been studied extensively~\cite{Espinosa:1991wt, FelixBeltran:2002tb, DiChiara:2008rg, Agashe:2011ia,Basak:2012bd, Delgado:2012sm,Delgado:2013zfa,Bandyopadhyay:2013lca,Kang:2013wm,deBlas:2013epa}; they offer richer phenomenology than singlet extensions, but they are also more constrained. Specifically, the neutral components of the triplets generically acquire vacuum expectation values (vev), causing tension with electroweak precision observables \cite{Khandker:2012zu,Englert:2013zpa}\footnote{Triplet extension which preserve custodial symmetry, such as the Supersymmetric Custodial Triplet Model, allow for large triplet vevs (and light scalars) without tension from electroweak precision observables \cite{Cort:2013foa, Garcia-Pepin:2014yfa,Delgado:2015aha}.}. Nonetheless, triplets offer appealing features when compared  with singlets, especially in the context of the DiracNMSSM: (i) more variety due to two possible hypercharge assignments, $Y=0$ or $Y=\pm1$, and (ii) triplets with hypercharge must be included in pairs for anomaly cancellation and can only have Dirac-type superpotential mass.

The rest of the paper is organized as follows. In Sec.~\ref{sec:model} we introduce the key superpotential interactions and give the correction to the Higgs mass for both the $Y=0$ and $Y=\pm1$ triplet models. Next, in Sec.~\ref{sec:ft} we analytically study the various sources of fine tuning, pinpointing the dependence of each term on the triplet parameters. This is followed up by a discussion of the precision electroweak  $\mathcal{T}$ parameter. From this discussion, it will be clear that the $Y = \pm 1$ model works better at raising the mass of the Higgs, avoiding fine tuning, and staying within the electroweak precision constraints. In Sec.~\ref{sec:Num} we perform a numerical study, focusing on the $Y = \pm 1$ scenario. As one of the primary differences between singlets and triplets is the existence of additional charged and potentially light fermions, in Sec.~\ref{sec:TripPheno} we review the phenomenology of `exotic' states, examining both direct production and indirect effects such as altered stop decays. Finally, conclusions are drawn in Sec.~\ref{sec:conclusions}.

%*********************Section***********************
\section{The Models}
\label{sec:model}
%*********************Section***********************

There are two signature features in the DiracNMSSM \cite{Lu:2013cta}, a Dirac mass term between two strictly different superfields, and the fact that only one of the two singlets couples to the Higgs doublets. The extension explored here, where a pair of triplets take the role of the singlets should maintain both properties. With this in mind, we define $\Sigma_1$ to be a $SU(2)_{L}$ triplet chiral superfield which couples to the Higgses in the superpotential, and a define a second triplet $\Sigma_{2}$, which does not. This is not the most general superpotential allowed by the symmetries of the model but we follow the setup of the original DiracNMSSM, in any case the choice is radiatively stable since superpotential couplings can not be generated via radiative corrections\footnote{One could also use a spurion analysis of an extra broken symmetry which would suppress the unwanted couplings~\cite{Lu:2013cta}.}.

With the inclusion of the triplets $\Sigma_{1,2}$ the superpotential is enlarged to
\begin{equation}
W=\mu H_{u}\cdot H_{d}+\mu_{\Sigma}\text{Tr}(\Sigma_{1}\cdot \Sigma_{2})+W_{H-\Sigma}+W_{\text{Yukawa}}
\label{eqn_SuperPotential}
\end{equation}
where the isospin product employs the convention $a\cdot b\equiv a_{i}\varepsilon_{ij}b_{j}$ with $\varepsilon_{21}=-\varepsilon_{12}=-1$.  The parameter $\mu_{\Sigma}$ is a supersymmetric Dirac mass for the triplets, $W_{H-\Sigma}$ couples $H_{u,d}$ with $\Sigma_{1}$ in a way specified by the hypercharge assignments of the triplets, and $W_{\text{Yukawa}}$ represents the standard MSSM Yukawa couplings.

We will analyze the cases $Y=0$ and $Y=\pm 1$ for the hypercharge of the triplets\footnote{These are the only possibilities that simultaneously permit a Dirac mass term and supply extra neutral scalars to raise $m_{h}^{2}$.}. When the triplets have hypercharge $Y=0$, they can couple to a combination of $H_u H_d$. This case should be seen as a simple extension of the singlet DiracNMSSM scenario, as the couplings take the same form up to factors of $\sqrt{2}$ coming from the normalization of the triplets. On the other hand, triplets with a hypercharge $Y=\pm1$ can only couple to $H_d^2$ or $H_u^2$. We examine the case where $H_u$ couples to the triplet but $H_d$ does not, since the latter will only generate an increased Higgs mass for the unphysical region of $\tan\beta<1$.  Both triplet scenarios contain charged scalars and fermions that are absent in the singlet DiracNMSSM. While potentially interesting at colliders, these extra states have minimal impact on the Higgs mass or fine tuning, so we will largely ignore them here. Comments on the phenomenology of the extra states can be found in Sec.~\ref{sec:TripPheno}.

%*********************subSection***********************
\subsection{$Y=0$ case}
%*********************subSection***********************

Triplets with hypercharge $Y=0$ couple to both $H_u$ and $H_d$ and are a simple extension to the singlet case studied in \cite{Lu:2013cta}. The superpotential is given by Eq.~(\ref{eqn_SuperPotential}) with
\begin{equation}
W_{H-\Sigma}=\lambda H_{d}\cdot \Sigma_{1}H_{u}.
\end{equation}
Forming the scalar potential, the superpotential terms are accompanied by the soft terms 
\begin{equation}
\Delta V_{\text{soft}} =m^2_{T}  \text{Tr}{|\Sigma_1|^2} + m^2_{\chi}  \text{Tr}{|\Sigma_2|^2}
+ \left( \lambda A_{\lambda} H_d \cdot \Sigma_1 H_u + \mu_{\Sigma} B_{\Sigma} \text{Tr}{(\Sigma_1 \cdot \Sigma_2)} + \text{h.c.}  \right),
\label{eq:vsoft}
\end{equation}
and the usual $SU(2)_{L}$ and $U(1)_{Y}$ $D$-terms. Here, $m_{T,\chi}$ are the triplet soft masses, $A_{\lambda}$ and $B_{\Sigma}$ are the trilinear and bilinear soft couplings respectively. While it is possible to give $Y=0$ triplets a non-Dirac supersymmetric mass, we ignore this possibility here as we are particularly interested in the effects of Dirac masses. Focusing on the \textit{CP} even scalar sector of the theory, the sole difference between the triplet and singlet MSSM extensions are factors of $\sqrt{2}$ coming from the normalization of the triplet. The full \textit{CP} even scalar potential for this scenario is shown in Appendix \ref{sec:appY0}.

Isospin triplets can potentially disrupt electroweak precision tests unless their vevs remain small. A simple way to mitigate the size of the triplet vevs is to take the scalar triplets to be heavier than the Higgses. In this limit, which we will assume throughout, the scalar triplets can be integrated out and are effectively replaced by combinations of lighter fields:
\begin{eqnarray}
\Sigma_{1,\text{neut}} \equiv T^0 & \rightarrow &\frac{\lambda}{\sqrt{2}} \frac{\mu (|H_u^0|^2 + |H_d^0|^2) - A_{\lambda} H_u^{0*} H_d^{0*}}{\mu_{\Sigma}^2 + m_T^2}+\mathcal{O}\left( \dfrac{1}{D_{T}^{2}},\dfrac{1}{D_{T}D_{\chi}},\dfrac{1}{D_{\chi}^{2}} \right) 
\label{eqn:defTy0} \\
\Sigma_{2,\text{neut}} \equiv \chi^0 & \rightarrow& \frac{\lambda \mu_{\Sigma}}{\sqrt{2}} \frac{H_u^0 H_d^0}{\mu_{\Sigma}^2 + m_{\chi}^2}++\mathcal{O}\left( \dfrac{1}{D_{T}^{2}},\dfrac{1}{D_{T}D_{\chi}},\dfrac{1}{D_{\chi}^{2}} \right).
\label{eqn:defCy0}
\end{eqnarray}
where $D_{T,\chi}\equiv \mu_{\Sigma}^{2}+m_{T,\chi}^{2}$.  The resulting effective potential for the Higgses can be found in Eq.~\eqref{eqn:VintoutY0}. From the effective potential, we can read off the modified tree-level \textit{CP}-even scalar mass matrices. Taking the decoupling limit for simplicity and adding the one-loop stop contribution to lightest tree-level mass eigenvalue, we find the Higgs mass:
\begin{eqnarray}
m_{h}^{2} &=& m_Z^2 \cos^2(2\beta) +  (\text{stop loops})+\frac{v^2 \lambda^2}{2}\sin^2(2\beta) \frac{m^2_{\chi}}{\mu^2_{\Sigma}+m^2_{\chi}} \nonumber \\
&&- \frac{v^2\lambda^2}{2} \frac{\left|2 \mu^* -A_{\lambda} \sin(2\beta)\right|^2}{\mu^2_{\Sigma}+m^2_{T}}.
\label{eqn_HiggsY0}
\end{eqnarray}

The expression above, with a positive (negative) piece that depends on the uncoupled (coupled) triplet soft mass is clearly reminiscent of the singlet DiracNMSSM, Eq. \eqref{eqn:singletModel}. As in the singlet case, the interplay between the two terms plays an important role in the fine tuning of the model.

%*********************subSection***********************
\subsection{$Y=\pm1$ case}
%*********************subSection***********************

Given that the superpotential should conserve hypercharge and be holomorphic, a supersymmetric mass term for a triplet with hypercharge $Y=1$ can only be included if there is a second triplet with $Y=-1$. Anomaly cancellation also rests on introducing hypercharge triplets in vector-like pairs. As in the $Y= 0$ scenario above, we assume $\Sigma_1$ is the triplet with superpotential couplings to the Higgses. Depending on its hypercharge $\Sigma_1$ will only be able to couple either to $H_u^2$ or $H_d^2$, which is distinct from the $Y=0$ setup. To get the largest impact from the triplet-Higgs coupling, we want it to couple as much as possible to the physical Higgs boson.  At large $\tan\beta$ and large $m_A$, the Higgs boson resides primarily in $H_u$, therefore we assign $Y = -1$ to $\Sigma_1$, permitting the interaction
\begin{equation}
W_{H-\Sigma}=\lambda H_{u}\cdot \Sigma_{1}H_{u}.
\end{equation}
The second triplet $\Sigma_2$ (now with hypercharge Y = 1) has no superpotential couplings. The soft terms are as in Eq. (\ref{eq:vsoft}) with the same modification to the $A_{\lambda}$ term as in the superpotential, and the complete \textit{CP} even scalar potential is given in Appendix \ref{sec:appY1}. 

When the triplet scalars are integrated out in this scenario, the neutral components are replaced by:
\begin{eqnarray}
\Sigma_{1,\text{neut}} \equiv T^0 & \rightarrow & \frac{\lambda \left(A_{\lambda} H_u^{0*}H_u^{0*}-2 \mu H_u^{0*} H_d^0 \right)}{\mu_{\Sigma}^2 + m_T^2}+\mathcal{O}\left( \frac{1}{D_\chi^2},\frac{1}{D_\chi D_T},\frac{1}{D_T^2} \right)
\label{eqn:defTY1} \\
\Sigma_{2,\text{neut}} \equiv \chi^0 & \rightarrow& \frac{-\lambda \mu_{\Sigma} H_u^0 H_u^0}{\mu_{\Sigma}^2 +m_{\chi}^2}+\mathcal{O}\left( \frac{1}{D_\chi^2},\frac{1}{D_\chi D_T},\frac{1}{D_T^2} \right).
\label{eqn:defCY1}
\end{eqnarray}
Working with the effective Higgs potential and proceeding as in the $Y = 0$ case, we find the decoupling-limit Higgs mass to be
\begin{eqnarray}
m_{h}^{2}
&=& m_Z^2 \cos^2(2\beta) +  (\text{stop loops}) + 4 v^2 \lambda^2 \sin^4(\beta)\left( \dfrac{ m_{\chi}^2}{\mu_{\Sigma}^2+m^2_{\chi}} \right) \nonumber\\
&&-\dfrac{v^2 \lambda^2 \sin^2{(2\beta)}}{\mu^2_{\Sigma} +m^2_{T}}\left|2\mu^* - A_{\lambda} \tan{(\beta)}\right|^2 .
\label{eqn_HiggsY1}
\end{eqnarray}

Comparing $m^2_h$ in the two models, Eqs. \eqref{eqn_HiggsY0} and \eqref{eqn_HiggsY1}, we see similar features. In both models there is a positive contribution to the Higgs mass proportional to $m^2_{\chi}/(\mu^2_{\Sigma}+m^2_{\chi})$. This is maximized when $m_\chi^2 \gg \mu_{\Sigma}^2$, and goes to zero when $m_{\chi}^2 \ll \mu_{\Sigma}^2$, so the Higgs mass is increased the most by decoupling the scalar part of $\Sigma_2$. In Section \ref{sec:ft} we will show that the decoupling of $m_{\chi}^{2}$ barely affects the fine tuning.

The amplitude and $\tan \beta$ dependence of the positive term is different for the $Y=0$ triplets and the $Y=\pm1$ triplets,
\begin{equation}
C_0(\beta)=\frac{v^2 \lambda^2}{2} \sin^2(2\beta)
\end{equation}
for $Y=0$ and 
\begin{equation}
C_1(\beta)= 4 v^2 \lambda^2 \sin^4 \beta.
 \end{equation}
for $Y=1$.
$C_0$ is maximized when $2\beta=\pi/2$, or $\tan\beta=1$. However, $C_1$ is maximal as $\beta \rightarrow \pi/2$, or $\tan\beta \rightarrow \infty$.
As the $\tan\beta$ dependence of $C_{1}$ aligns with that of the MSSM, the size of the triplet contributions to the Higgs mass do not need to be as large, leading to smaller values of $\lambda$ in the $Y=\pm1$ model.

Equations \eqref{eqn_HiggsY0} and \eqref{eqn_HiggsY1} also have a term which acts to lower $m^2_h$. The negative terms depend on the mass of $\Sigma_1$, the triplet which couples to the doublets. A large soft mass for $\Sigma_1$ decreases the absolute value of the negative term, raising the Higgs mass. However, $m^2_T$ also enters into the radiative corrections of the Higgs soft masses, so the $m_T$ value that minimizes the fine tuning is less clear cut and is best tackled numerically.

Both of the negative terms also contain a factor which depends on the difference between $\mu$ and $A_{\lambda}$, $\left| 2 \mu^* - A_{\lambda} \sin(2\beta) \right| ^2$ for the $Y=0$ case and $\left| 2 \mu^* - A_{\lambda} \tan \beta \right|^2$ for $Y=\pm1$ respectively. The same expressions appear in the effective triplet vevs, Eq.(\ref{eqn:defTy0}, \ref{eqn:defCy0}) or Eq.(\ref{eqn:defTY1},\ref{eqn:defCY1}) after the Higgs doublets acquire vacuum expectation values. The $\mathcal{T}$ parameter is tightly constrained by precision electroweak measurements, however, the fact that the same expressions appear in the Higgs mass and the triplet effective vevs implies that regions with the smallest negative contribution to the Higgs mass are also the regions with the smallest $\mathcal{T}$ parameter. 

Having shown how the Higgs mass is altered in the two Dirac Triplet scenarios and identified key parameters, we now move on to study the fine tuning.
%*********************Section***********************
\section{Fine tuning calculations and $\mathcal T$ parameter}
\label{sec:ft}
%*********************Section***********************

Equations \eqref{eqn_HiggsY0} and \eqref{eqn_HiggsY1} show that decoupling the soft mass of $\Sigma_2$ leads to a maximal increase in the Higgs mass. Ordinarily, the introduction of large scalar masses to correct the Higgs mass increases the fine tuning. In the next subsection we show that this is not the case for this model; the fact that $\Sigma_2$ does not couple to the doublets allows it to be decoupled with small effects on the fine tuning,  as in the original DiracNMSSM model. Beyond the fine tuning of the Higgs mass, triplet models are also constrained by the $\mathcal{T}$ parameter, which we examine more closely in Section \ref{subsec:T}.

%*********************Section***********************
\subsection{Fine tuning of $m^2_{H_u}$}
\label{subsec:MH2}
%*********************Section***********************
We adopt the definition of fine tuning of \cite{Lu:2013cta},
\begin{equation}
\Delta = \frac{2}{m^2_h} \text{max}\left(m^2_{H_u}, m^2_{H_d}, \frac{d m^2_{H_u}}{d\log{(u)}} L, \frac{d m^2_{H_d}}{d\log{(u)}} L, \delta m_{H_{u}^{0}}^{2}, \mu B_{\mu,\text{eff}} \right)
\label{eqn:fine-tuning}
\end{equation}
where $L\equiv \log(\Lambda/m_{\widetilde{t}})$ accounts for the running to the SUSY breaking scale, $\log(u)$ is the running scale and $\delta m_{H_{u}^{0}}^{2}$ is the one-loop finite threshold correction from the triplets; following \cite{Lu:2013cta}, we set $L=6$. Although we use the same definition for $\Delta$ that was used for the singlet model, we expect the triplet case to be slightly different due to larger triplet-Higgs couplings (coming from the normalization of the triplets) and the different hypercharge possibilities. Putting all of the components together and taking the maximum contribution is best done numerically. However, before launching into numerics, in this section we examine each of the different components of $\Delta$ to get a better feeling for their relative importance and to see how they depend on the triplet parameters.

The first entries in $\Delta$ are $m^2_{H_u}$ and $m^2_{H_d}$, the tree-level soft masses for the Higgs doublets. These are not free parameters, rather they are set by the requirement that electroweak symmetry is broken at the minimum of the scalar potential (see Eq.~\eqref{eqn:mincondHu} and \eqref{eqn:mincondHd} for $Y=0$ and \eqref{eqn:mincondHuY1} and \eqref{eqn:mincondHdY1} for $Y=\pm1$). In solving the minimization conditions, $m^2_{H_u}$ and $m^2_{H_d}$ inherit a complicated dependence on the triplet parameters that is difficult to generalize. As these entries are typically subdominant in $\Delta$, we do not attempt to tease out the triplet parameter dependence analytically.

The next components of $\Delta$ are $\frac{d m^2_{H_u}}{d\log{(u)}} L, \frac{d m^2_{H_d}}{d\log{(u)}} L$, the radiative corrections to the Higgs soft masses. While nominally one-loop effects, these radiative pieces have the potential to be important because they depend quadratically on the masses of heavy particles (stops, triplets, etc.) -- objects that do not appear or are subdominant in the tree level Higgs potential. Additionally, the radiative effects are enhanced by $L$, the logarithm that encapsulates the running of soft masses down from the supersymmetry mediation scale. As a result, these radiative pieces are often the largest component of $\Delta$. To see how the triplet parameters enter, we need the renormalization group equations (RGE) governing the evolution of $m^2_{H_u}, m^2_{H_d}$: 
\begin{equation}
(Y=0) ~~~ \left\{
 \begin{matrix}
16 \pi^2 \frac{d m^2_{H_u}}{dt} \supset 6 h_t^2 \left(m^2_{Q_3} + m^2_{U_3} + m^2_{H_u} \right) + 6 \lambda^2 \left(m^2_{H_u} + m^2_{H_d} + m^2_{T} + A_{\lambda}^2 \right)\\
16 \pi^2 \frac{d m^2_{H_d}}{dt} \supset 6 h_b^2 \left(m^2_{Q_3} + m^2_{D_3} + m^2_{H_d} \right) + 6 \lambda^2 \left(m^2_{H_u} + m^2_{H_d} + m^2_{T} + A_{\lambda}^2 \right)
\end{matrix} \right.
\label{eqn:HuRGEY0}
\end{equation}
and
\begin{equation}
(Y=\pm1) ~~~  \left\{
 \begin{matrix}
16\pi^2 \frac{dm^2_{H_u}}{dt} \supset 6 h_t^2 \left( m^2_{Q_3} + m^2_{U_3} + m^2_{H_u} \right) + 6 \lambda^2 \left( 2 m^2_{H_u} + m^2_{T} + A_{\lambda}^2 \right) \\
16\pi^2 \frac{d m^2_{H_d}}{dt}\supset 6 h_b^2 \left(m^2_{Q_3} + m^2_{D_3} + m^2_{H_u}  \right)
\end{matrix} \right. .
\label{eqn:HdRGEY1}
\end{equation}
The large top Yukawa, $h_t$ and the dependence on the stop masses needed in the MSSM to raise the Higgs mass are what drives the fine tuning.  In the triplet scenario, the extra contributions to the (tree level) Higgs mass from the triplets permits lighter stops and allows for a less tuned model.

The key difference between the DiracNMSSM and the traditional NMSSM is that the mass of the uncoupled state does not feed into the Higgs soft masses at loop level. This same behavior is reproduced in Eq~(\ref{eqn:HuRGEY0}) nor (\ref{eqn:HdRGEY1}) , neither of which depends on $m_{\chi}$, the mass of  $\Sigma_2$. As a result,  large $m_{\chi}$ -- and thereby large positive contributions to the Higgs mass -- are permitted without giving rise to fine tuning. The soft mass of $\Sigma_1$ and the trilinear soft term $A_{\lambda}$ enter into the running of $m^2_{H_u}, m^2_{H_d}$, so in principle large values for them would increase $\Delta$. However, both $m_T^2$ and $A_{\lambda}^2$ enter into the beta functions multiplied by $\lambda^{2}$, hence a smaller $\lambda$ would permit these two quantities to take moderate values without dominating the fine tuning.

Following the radiative piece in $\Delta$ is the threshold correction $\delta m_{H_{u}^{0}}^{2}$, the finite contribution to $m^2_{H_u}$ that emerges when heavy fields are integrated out. The threshold terms are important as they are the only place where the soft mass of the uncoupled triplet $m_{\chi}^{2}$ (or the non-coupling singlet, in the model of Ref.~\cite{Lu:2013cta}) enters into the fine tuning. The threshold corrections, presented in full in Appendix \ref{sec:AppFTC}, depend on the soft masses of both triplets.  However, since $m_T$ also appears in the (log-enhanced) RGE part of the tuning discussed above, keeping $m_T$ small minimizes the tuning. With $m_T$ kept small, the threshold correction is well approximated by the $\Sigma_2$ piece alone:
\begin{equation}
\begin{aligned}
(Y=0):~~~~ \delta m_{H_u^0}^2 & \simeq \frac{3}{2}\frac{\lambda^2 \mu_{\Sigma}^2}{16\pi^2} \log\frac{m^2_{\chi} + \mu_{\Sigma}^2}{\mu_{\Sigma}^2} \text{ and} \\
(Y=\pm1):~~~~  \delta m_{H_u^0}^2 & \simeq 6 \frac{ \lambda^2 \mu_{\Sigma}^2}{16 \pi^2}  \log \frac{m^2_{\chi} + \mu_{\Sigma}^2}{\mu_{\Sigma}^2}.
\label{eqn:thresholdCorrection}
\end{aligned}
\end{equation}
If $\mu_{\Sigma}^2 \gtrsim m_{\chi}^2$, there is little fine tuning from the threshold correction. We saw in Sec.~\ref{sec:model} that the most interesting parameter space -- where the triplet contribution to the Higgs mass is large and positive -- occurs when $m_{\chi}^2 \gg \mu^2_{\Sigma}$. For this hierarchy of parameters,  the threshold contribution can be non-negligible, though only when $\mu^2_{\Sigma}$ is large (compared to $m_h$) as well. 

The final component of $\Delta$ is the dependence on $\mu$ and $B_{\mu}$.  For the triplet scenario with hypercharge, this component of the tuning is identical to the MSSM. Triplets without hypercharge are slightly more complex, since the effective triplet vevs shift $\mu$ and $B_{\mu}$ from their MSSM values. The shifted values are given by
\begin{align}
\mu_{eff} &= \mu - \frac{\sqrt{2}}{2}\lambda \left\langle T^0 \right \rangle \text{ and} \label{eqn:muEff} \\ 
\mu B_{\mu,\text{eff}} &= \mu B_{\mu} - \frac{\lambda}{\sqrt{2}} \left(A_{\lambda} \left\langle T^0 \right \rangle + \mu_{\Sigma} \left \langle \chi^0 	\right \rangle \right ). \label{eqn:BEff}
\end{align}
Though not usually the dominant component in $\Delta$, these contributions to the fine tuning measure are inevitable as $\mu$ and  $B_{\mu}$ enter directly into the tree-level mass matrix of the Higgs.

After considering the individual components of the fine tuning measure,  we are now ready for a full numerical study of the tuning over a range of triplet parameters. Before doing so, we first examine how the $\mathcal{T}$ parameter constrains the available parameter space.

%*********************SubSection***********************
\subsection{Constraints from the $\mathcal{T}$ parameter}
\label{subsec:T}
%*********************SubSection***********************

Electroweak scalar triplets that acquire vacuum expectation values notoriously spoil the relation between $m_W$ and $m_Z$. This mass ratio is more commonly expressed as the $\mathcal T$ parameter
\begin{equation}
\alpha \mathcal{T} = \frac{m_W^2}{m_Z^2 \cos^2 \theta_W}-1.
\end{equation}
The authors of \cite{Baak:2012kk,Baak:2014ora} used data from $Z$ pole measurements \cite{ALEPH:2005ab}, the running quark masses \cite{Beringer:1900zz}, the five-quark hadronic vanuum polarization contribution to $\alpha\left(M_Z^2\right)$, $\Delta\alpha_{\text{had}^{(5)}} \left(M_Z^2\right)$ \cite{Davier:2010nc}, the mass and width of the $W$ \cite{Beringer:1900zz}, top quark mass \cite{ATLAS:2014wva}, and Higgs mass measurements \cite{Aad:2014aba,CMS:2014ega} to preform a global fit of electroweak data. A value of $\mathcal{T} = 0.09\pm 0.13$ gives the best fit of the data if all of the oblique parameters are allowed to float\footnote{If the $U$ parameter is fixed to $U=0$, the best fit is $\mathcal{T}=0.10\pm0.07$.}.
Forcing the (tree-level) triplet contributions to the $\mathcal T$ parameter to lie within the 1-$\sigma$ uncertainty, we can derive a bound on the triplet model parameters. 
In an effective theory where we have integrated out the triplets, there are no triplet fields around to get vevs, but the $\mathcal T$ contributions are still present in the form of higher dimensional operators.
Specifically, after integrating out the triplets, the kinetic term for the $\Sigma_i$ becomes (schematically)
\begin{equation}
\left| D_{\mu} \Sigma_i \right|^2 \xrightarrow[\text{integrated out}]{\Sigma} \frac{1}{\Lambda^2} \left| H D_{\mu} H \right|^2,
\end{equation}
which, once the Higgses are set to their vevs, contributes differently to the $W$ and $Z$ mass. The operator is intentionally left vague, as the actual combinations of the $H_u$ and $H_d$ and the mass scale $\Lambda$ are different for each triplet.

For the triplets with $Y=0$, this operator contributes to $\mathcal{T}$ by
\begin{equation}
\mathcal{T}_{Y=0}=\dfrac{1}{\alpha} ~ \dfrac{4\bigl( \langle \chi^{0}\rangle^{2}+\langle T^{0}\rangle^{2} \bigr)}{v^{2}-4\bigl( \langle \chi^{0}\rangle^{2}+\langle T^{0}\rangle^{2} \bigr)}
\label{eqn:TparamY0}
\end{equation}
where $\langle T^0 \rangle$ and $\langle \chi^0 \rangle$ are the values of equation \eqref{eqn:defTy0} and \eqref{eqn:defCy0} after the doublets have developed vevs  -- what we dub `effective vevs' for the triplets. The effective vevs are approximately given by
\begin{equation}
\left\langle T^0 \right\rangle_{Y=0} \approx \dfrac{v^2 \lambda}{2\sqrt{2}} \dfrac{2 \mu^* - A_{\lambda}\sin(2\beta)}{\mu^2_{\Sigma}+m^2_{T}} \text{ and}
 ~~~~
\left\langle \chi^0 \right \rangle_{Y=0}  \approx -\dfrac{v^2 \lambda}{2\sqrt{2}} \dfrac{\mu_{\Sigma} \sin(2\beta)}{\mu^2_{\Sigma} + m^2_{\chi}},
\label{eqn_vev0}
\end{equation}
up to higher order terms in $1/(\mu_{\Sigma}^{2}+m_{T,\chi}^{2})$. For the case with hypercharge, the $\mathcal{T}$ parameter takes the form
\begin{equation}
\mathcal{T}_{Y=\pm1} = -\frac{1}{\alpha}~ \frac{2 \left(\langle \chi^0 \rangle^2 + \langle T^0 \rangle^2 \right) }{v^2}
\label{eqn_Y1}
\end{equation}
with $\langle T^0 \rangle$ and $\langle \chi^0\rangle$ now coming from Eq. \eqref{eqn:defTY1} and \eqref{eqn:defCY1} once the doublets acquired the vevs, 
\begin{equation}
\left\langle T^0 \right \rangle_{Y=-1}  \approx -\dfrac{v^2\lambda}{2} \dfrac{ \sin(2\beta) \left(  2 \mu^{*} -A_{\lambda} \tan{(\beta)}  \right)}{ \mu^2_{\Sigma}+m^2_{T} }\text{ and}
 ~~~~
\left\langle \chi^0 \right\rangle_{Y=1} \approx v^2 \dfrac{ - \lambda \mu_{\Sigma} \sin^2{(\beta)}}{ \mu^2_{\Sigma}+m^2_{\chi} }.
\label{eqn_vev1}
\end{equation}

Inspecting these equations, we can identify several parameter combinations that dictate the size of the $\mathcal{T}$ parameter. 

\begin{itemize}

	\item $m_{\chi}^{2}$: The effective vev $\bra \chi^0\ket \rightarrow 0$ in the limit of large $m_{\chi}$. In order to effectively raise the Higgs mass, we want $m_{\chi}^2 \gg \mu^2_{\Sigma}$. Large $m_{\chi}$ also does not add to the fine tuning (see previous subsection), so large $m_{\chi}$ is preferred for both the fine tuning and the $\mathcal{T}$ parameter.

	\item$m_{T}^{2}$: Similarly, the effective vev $\bra T^0 \ket \rightarrow 0$ in the limit of large $m_{T}$. A large value for $m_{T}$ also reduces the negative term in the Higgs mass squared equations. However, $m^2_T$ enters into the tuning from the RGE running terms and can quickly dominate the fine tuning.
		
	\item $\mu_{\Sigma}$: Both $\bra \chi^0 \ket \rightarrow 0$ and $\bra T^0 \ket\ \rightarrow 0$ for large $\mu_{\Sigma}$. This is not desired as it decreases the triplet contribution to the Higgs mass and removes any interesting phenomenology of extra light states. 
	
	\item $\lambda$: The $\mathcal{T}$ parameter goes as $\lambda^2$. The fact that the $Y=\pm1$ model can easily get the correct Higgs mass for lower values of $\lambda$ implies that the model with hypercharge will not be as constrained by the $\mathcal{T}$ parameter for fixed stop masses.

	\item$\mu$ \textit{and} $A_{\lambda}$: One could also have a cancelation between the $\mu$ and $A_{\lambda}$ terms. This would be a cancellation between a supersymmetric term and a soft term, which is in itself unnatural.

	\item $\tan \beta$: The triplets with hypercharge $Y=\pm1$ have an extra dependence on $\sin(2\beta)$ in $\bra T^0\ket$. At large values of $\tan\beta$, this goes to 0. Large values of $\tan\beta$ were already preferred for $Y=\pm1$ in order to raise the Higgs mass as much as possible. The $Y=0$ model is not as lucky. 
	
\end{itemize} 

Considering these points, in particular the $\lambda$ and $\tan\beta$ dependence, it is clear that the $\mathcal{T}$ parameter is more constraining on the $Y=0$ model. In addition, for fixed triplet-Higgs coupling $\lambda$, the triplet contribution to the Higgs mass in the $Y = 0$  model is smaller than in the singlet DiracNMSSM scenario because of the $\sqrt 2$ factor in the normalization of the neutral components. As this scenario suffers in fine tuning and the $\mathcal{T}$ parameter without the promise of interesting phenomena, we choose to ignore the $Y=0$ Dirac triplet model for the rest of the paper and focus our numerical and phenomenological study on $Y \ne 0$.

Lastly, we point out that $\bra T^0 \ket$ and $\bra \chi^0\ket$ contribute to the $\mathcal{T}$ parameter at tree level, and to order $\lambda^2$. To be consistent, we have also calculated the one-loop fermionic contributions to the $\mathcal{T}$ parameter to order $\lambda^2$. Because the triplet fermions are Dirac particles, and the mixing to order $\lambda^2$ keeps the entire triplet representation the same mass, there is no contribution to the $\mathcal{T}$ parameter at order $\lambda^2$.

%*********************Section***********************
\section{Numerical study: $Y = \pm 1$}
\label{sec:Num}
%*********************Section***********************

The analytical expressions of the last section allowed us to determine the overall scheme needed to minimize fine tuning and yet maximize the triplet contributions to the Higgs mass. Focusing entirely on the $Y = \pm 1$ scenario, the preferred regions are large $\tan\beta$, large $m_{\chi}$, and small values for $m_{T}$ and the stop masses. The coupling $\lambda$ needs to be large enough to raise the Higgs mass without being so large as to induce large triplet vevs. While there are multiple free parameters at hand, we wish to keep our numerical analysis both detailed and manageable. For this reason, we limit the parameters we vary to two scans, one over $\lambda$ and $m_T$ and the other over $\mu_{\Sigma}$ and $m_{\chi}$. The other parameters are fixed to benchmark values shown in Table.~\ref{tab_BenchmarkFT}.\\

\begin{table}[h!]
\centering
\begin{tabular}{lcc}
\hline
\hline
$\tan{\beta} = 10$& $m_A = 300\gev$ & $A_t=0$ \\
$\mu = 250\gev$ & $B_{\Sigma} = 100\gev$ & $A_{\lambda}=0$\\
\hline
\hline
\end{tabular}
\caption{Benchmark parameter values for the calculation of the fine tuning variables for the $Y=\pm1$ model. For simplicity, the gaugino masses and all squark/slepton masses other than the stop are assumed to be decoupled.}
\label{tab_BenchmarkFT}
\end{table}

The values for the fixed parameters in Table~\ref{tab_BenchmarkFT} are motivated by several considerations. First, since the Higgs mass contribution, fine tuning, and $\mathcal T$ parameter are improved at large $\tan{\beta}$, we select $\tan\beta = 10$ as a representative value.  Next, the scalar masses $m_A$ and $B_{\Sigma}$ play little role in our results, so they are good parameters to fix. The mass $m_A$ enters into the Higgs mass matrix, however as we always assume the decoupling limit it has little effect (so long as the value we choose is large enough to justify the decoupling limit). Similarly, the soft parameter $B_{\Sigma}$ mixes the scalars from $\Sigma_1$ and $\Sigma_2$. This mixing does not change our results, but complicates the translation between the scalar mass eigenstates and the Lagrangian parameters. Therefore we select a small $B_{\Sigma}$ for simplicity.

The effective vev $\bra T^0 \ket$ (and therefore the $\mathcal{T}$ parameter) depend on the difference between $\mu$ and $A_{\lambda}$, however, this term is suppressed at large values of $\tan\beta$. Varying $A_{\lambda}$ over a moderate range of values, we find the fine tuning does not change much. Therefore, we set $A_{\lambda}$ to 0 (together with $A_{t}$ for consistency), a choice that fits well within gauge mediated SUSY breaking scenarios \cite{Dine:1981gu, Nappi:1982hm,AlvarezGaume:1981wy, Dine:1993yw,Dine:1994vc,Dine:1995ag}.

The last parameter we fix is $\mu$. Since we have decoupled/ignored the wino, the chargino mass is set by $\mu$, thus the existing LEP2 bound~\cite{lepii} on charginos sets a lower bound of $\mu \gtrsim 100\,\gev$. High $\mu$ values are also disfavored by fine tuning, so we therefore pick an intermediate value of $\mu = 250\,\gev$ for our benchmark. The contribution to the tuning for this choice $\Delta(\mu)=8.47$; as this value is independent of the rest of the spectrum, $\Delta(\mu)$ should be regarded of as the minimum tuning possible according to our measure. From the fine tuning perspective alone, a value of $\mu$ closer to the LEP2 bound would be better. However, as we will detail in section \ref{sec:TripPheno}, $\mu$ also plays a role in stop phenomenology.
  
To study the fine tuning, we scan over the remaining triplet parameters, the coupling $\lambda$, the Dirac mass, $\mu_{\Sigma}$, and the soft masses, $m_{\chi}$ and $m_T$. Once values for these are chosen, the triplet contribution to the Higgs mass is known (see Eq.\eqref{eqn_HiggsY1}) and the stops are the only part left to enforce $m_h=125\gev$. As the stop contribution to the Higgs mass depends on the masses of both stops, we must make some assumptions in order to extract the values. We study two different assumptions:
\begin{enumerate}
	\item Left and right-handed stop have the same mass. ($m_{\tilde{Q}_3}=m_{\tilde{u}^c_3}$)
	\item The right-handed stop is used to set the Higgs mass while the left-handed stop is set to 800 GeV, which is above the most stringent LHC limits \cite{Aad:2012xqa,Aad:2014qaa,Aad:2012uu,Aad:2014nra,Aad:2014bva, Aad:2012ywa, Chatrchyan:2012lia, Chatrchyan:2013xna, Chatrchyan:2014lfa, CMS:2014yma, CMS:2014wsa}. 
\end{enumerate}
Next, we use SuSpect2 \cite{Djouadi:2002ze} to find the mass of the Higgs in the MSSM for the benchmark values and a given set of stop masses. The final Higgs mass squared is then the result of adding the MSSM part and the triplet contribution in quadrature. 
\begin{equation}
m_h^2 \equiv (125.5\gev)^2 = m_h^2 (\text{MSSM}) + m_h^2 (\text{Triplet}).
\end{equation}
We vary the value of the stop mass until this relationship is achieved. Then, once the stop mass is known, we can calculate the fine tuning defined in Eq.~\eqref{eqn:fine-tuning}.

\begin{figure}[t]
\begin{center}
\begin{subfigure}[t]{\textwidth}
\includegraphics[width=0.45\linewidth]{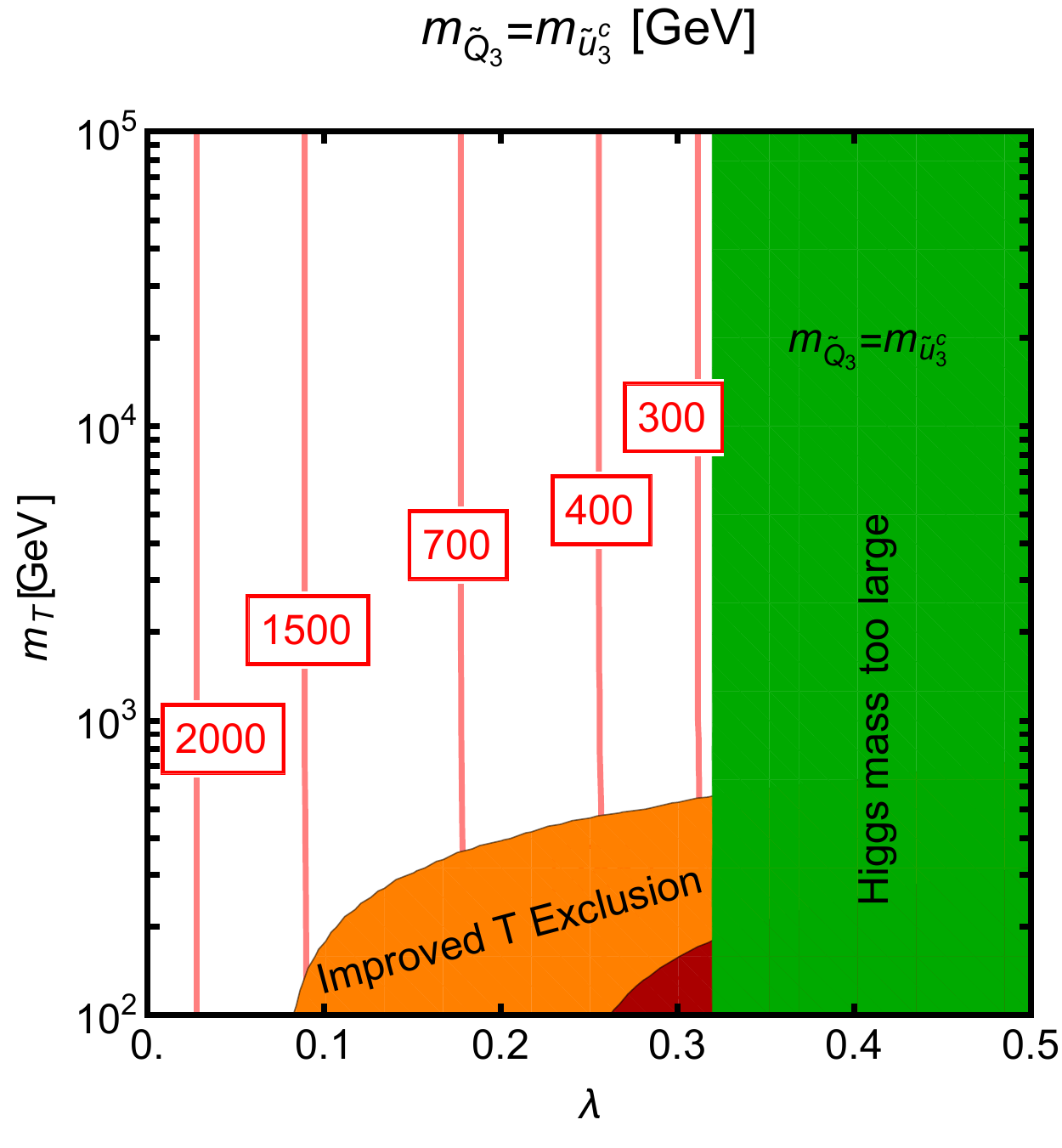}
\includegraphics[width=0.45\linewidth]{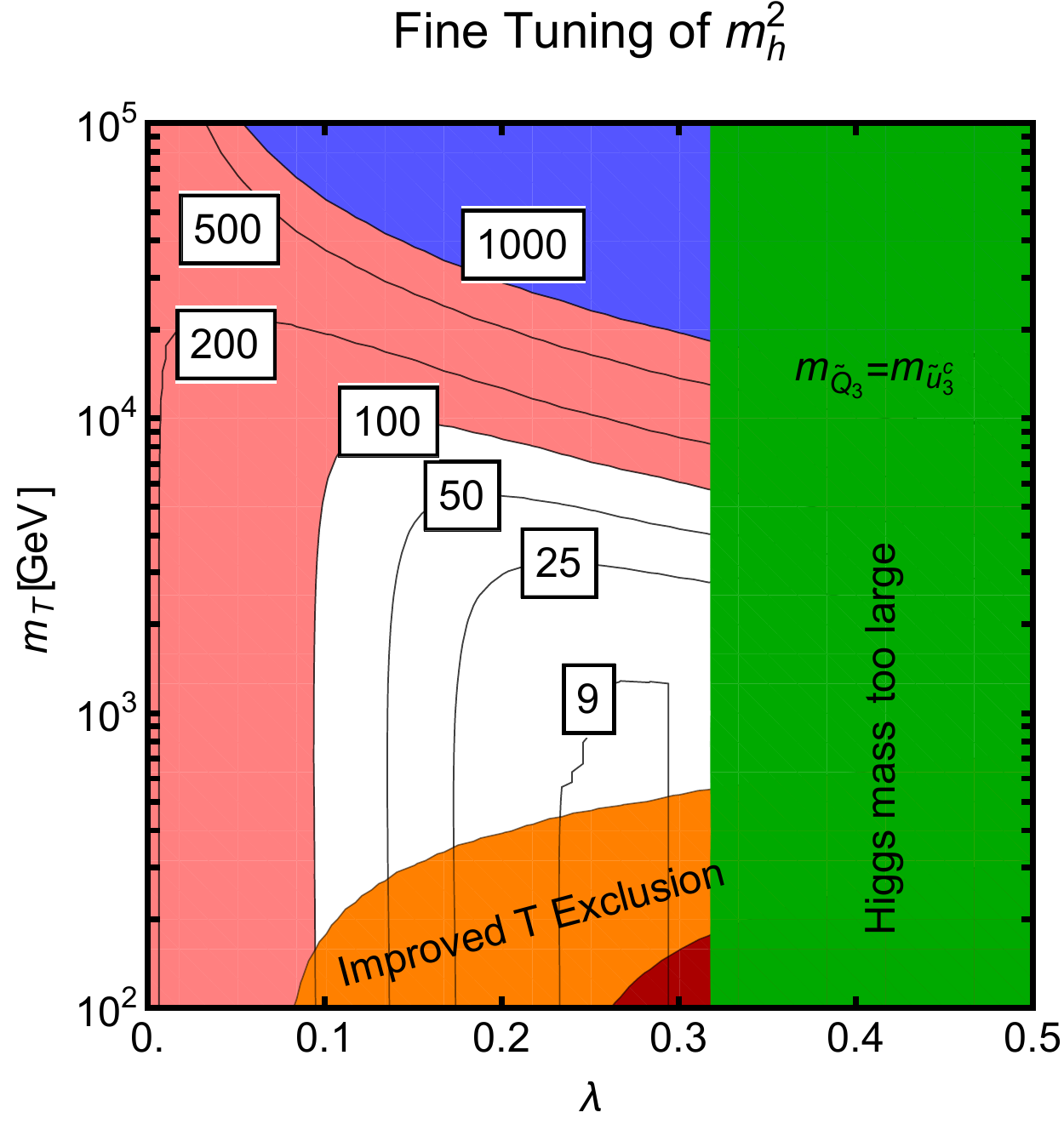}
\caption{Left and right-handed stop masses are equal.}
\label{fig:finetuningTB10BothEqualLambda}
\end{subfigure}
\begin{subfigure}[t]{\textwidth}
\includegraphics[width=0.45\linewidth]{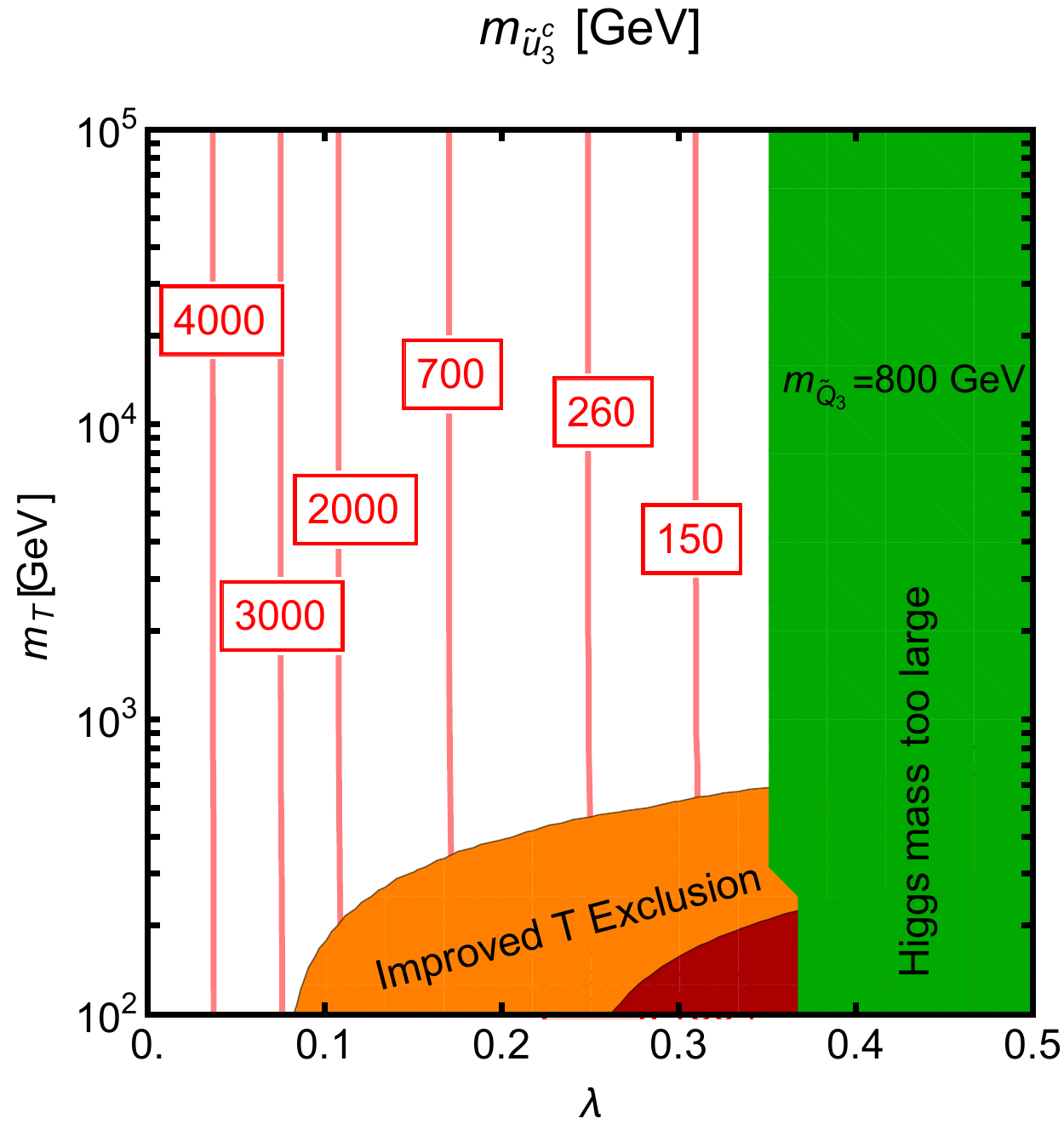}
\includegraphics[width=0.45\linewidth]{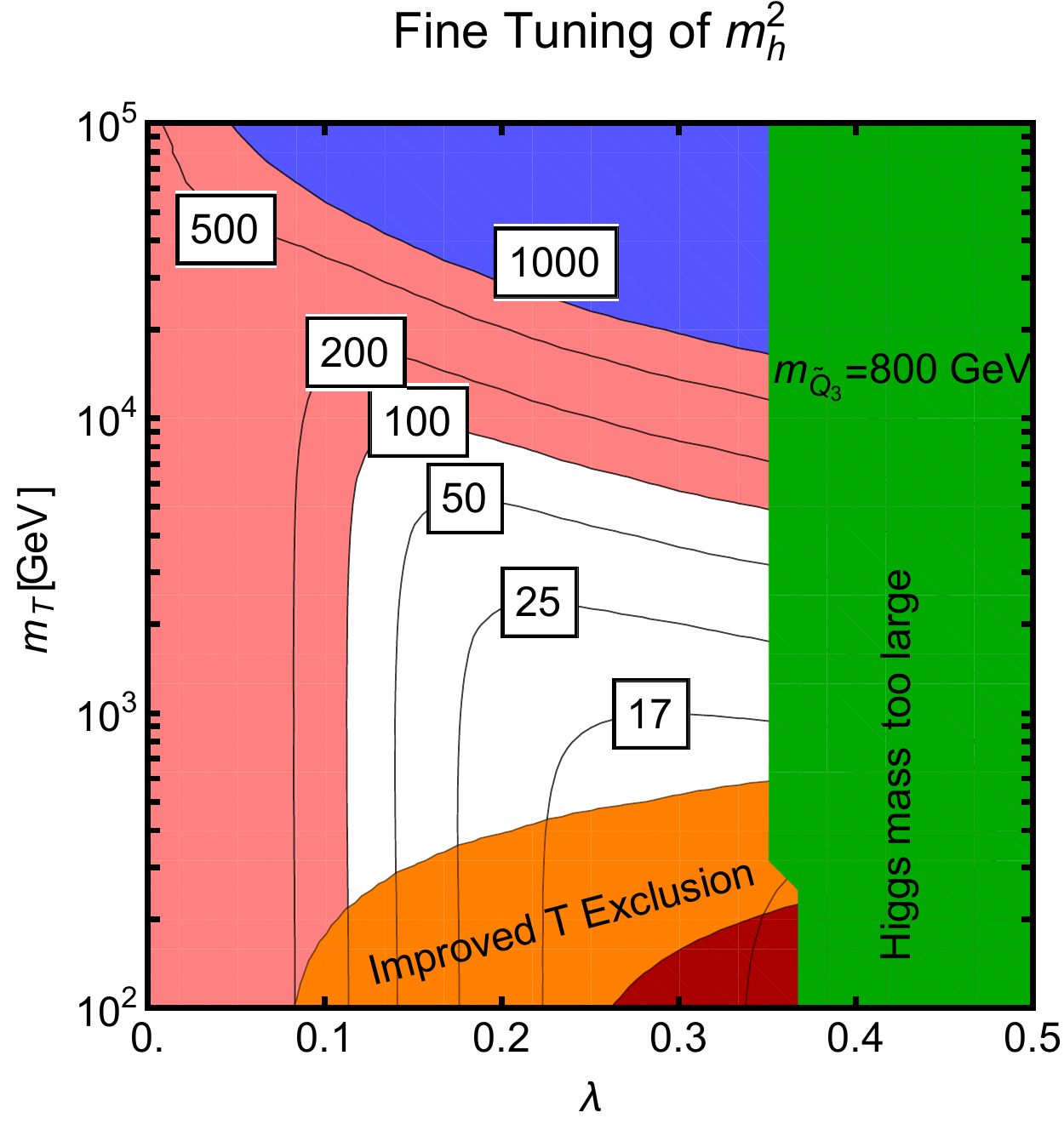}
\caption{Left-handed stop set to $800\gev$ and the right-handed stop sets the Higgs mass.}
\label{fig:finetuningTB10ChangeRightLambda}
\end{subfigure}
\caption{The left panels show contours of the stop soft mass needed in order to raise the Higgs mass to he observed value when $\mu_{\Sigma}=300\gev$, $m_{\chi}=10\tev$ and $\tan\beta=10$. In (\subref{fig:finetuningTB10BothEqualLambda}) both stops have the same mass while (\subref{fig:finetuningTB10ChangeRightLambda}) only changes the right-handed soft mass and keeps the left-handed stop at $800\gev$. The right panels show the corresponding contours of fine tuning. The dark red region marks where the vevs of the triplets cause too-large  contributions to the $\mathcal{T}$ parameter. The orange region supposes an improvement in the measured $\mathcal{T}$ parameter by an order of magnitude.}
\label{fig:ftLambda}
\end{center}
\end{figure}

Knowing that the triplet contribution to the Higgs mass is largest when $m_{\chi} \gg \mu_{\Sigma}$, we first choose to fix
\begin{equation}
m_{\chi}=10 \tev ~~~\text{ and } ~~~ \mu_{\Sigma}=300\gev
\end{equation}
and scan over value of $\lambda$ and $m_T$. The left panels of Fig.~\ref{fig:ftLambda} show the values of the stop soft masses that are needed in order to set the correct Higgs mass; in Fig.~\ref{fig:finetuningTB10BothEqualLambda}, both stop soft masses are equal, while in Fig.~\ref{fig:finetuningTB10ChangeRightLambda} the left-handed soft mass is fixed at $800\gev$ and the right-handed soft mass is indicated by the contours. The triplets do not affect the Higgs mass in the MSSM limit that $\lambda\rightarrow0$, so very large stop masses are needed. As $\lambda$ is increased from zero, the necessary stop mass decreases. If $\lambda \gtrsim 0.35$, the triplet $F$-terms generate a Higgs mass that is alway greater than observed value. These regions are marked in green in the figures. The soft mass $m_T$ only affect the mass of the Higgs through the negative term in Eq.~(\ref{eqn_HiggsY1}). For large values of $\tan\beta$, this term is negligible. 

The fine tuning is calculated at each point once the stop masses have been obtained. Contours of $\Delta$ are shown in the right panels of Fig.~\ref{fig:ftLambda}. The white, pink, and blue regions represent a fine tuning of $\Delta \le 100$, $100 < \Delta \le 1000$, and $\Delta > 1000$ respectively. The RGE running part of the fine tuning measure is dominant and depends on  $h_t^2 (m^2_{Q_3} + m^2_{U_3})$ and $\lambda^2 m_T^2$. Increasing $\lambda$ lowers the stop masses, decreasing the fine tuning until $\lambda^2 m_T^2$ is comparable to $h_t^2 (m^2_{Q_3} + m^2_{U_3})$. As such, a small value of the soft mass is preferred for fine tuning, although the $\mathcal{T}$ parameter can cause issues if $m_T$ is too light. 

At each point in the scan we calculate the effective triplet vevs and their contribution to the $\mathcal{T}$ parameter. The red regions show where the triplet contributions to $\mathcal{T}$ are larger than the 0.13 1-$\sigma$ uncertainty \cite{Baak:2014ora}. We also mark in orange what could be excluded by a new precision study of the $Z$-pole if the uncertainty on the $\mathcal{T}$ parameter were decreased by an order of magnitude. Fig.~\ref{fig:ftLambda} has the soft mass of $\Sigma_2$ decoupled ($m_{\chi}=10\tev$), so $\bra \chi^0\ket$ is negligible and $\mathcal{T}$ is only affected by $\bra T^0 \ket$. The large value of  $\tan\beta$ suppresses $\bra T^0 \ket$ so the current $\mathcal{T}$ bounds can only exclude $m_T < 200 \gev$ at the largest allowed values of $\lambda$. An improved measurement brings the exclusion to values of $\lambda$ as low as 0.1 and soft masses as large as $500\gev$. The vev $\bra T^0 \ket$ is proportional to $1/(\mu_{\Sigma}^2+m_T^2)$, so the reach of this exclusion region is strongly dependent on the value of $\mu_{\Sigma}$ as well, which has been kept fixed up to this point.

\begin{figure}[t]
\begin{center}
\begin{subfigure}[t]{\textwidth}
\includegraphics[width=0.45\linewidth]{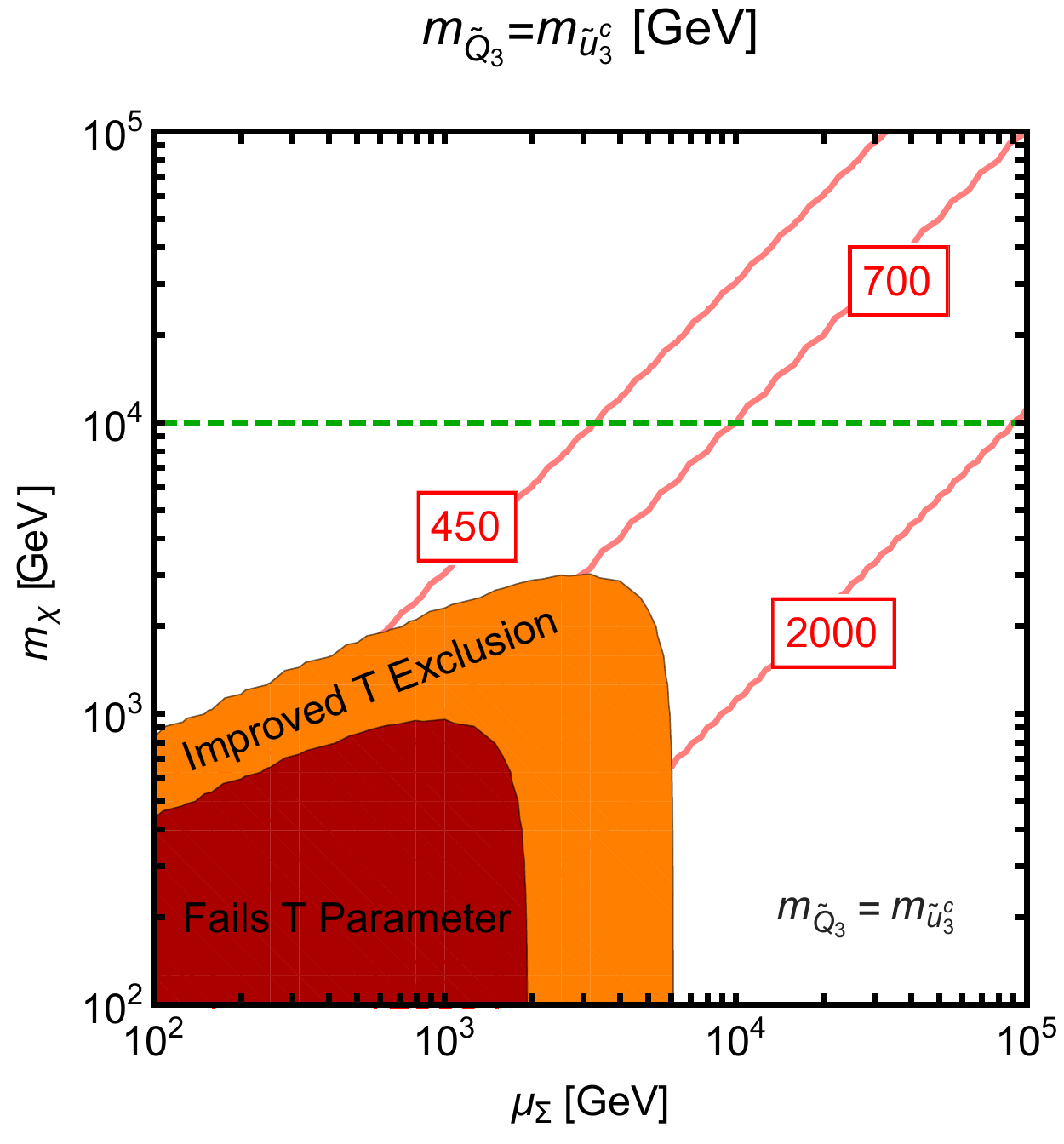}
\includegraphics[width=0.45\linewidth]{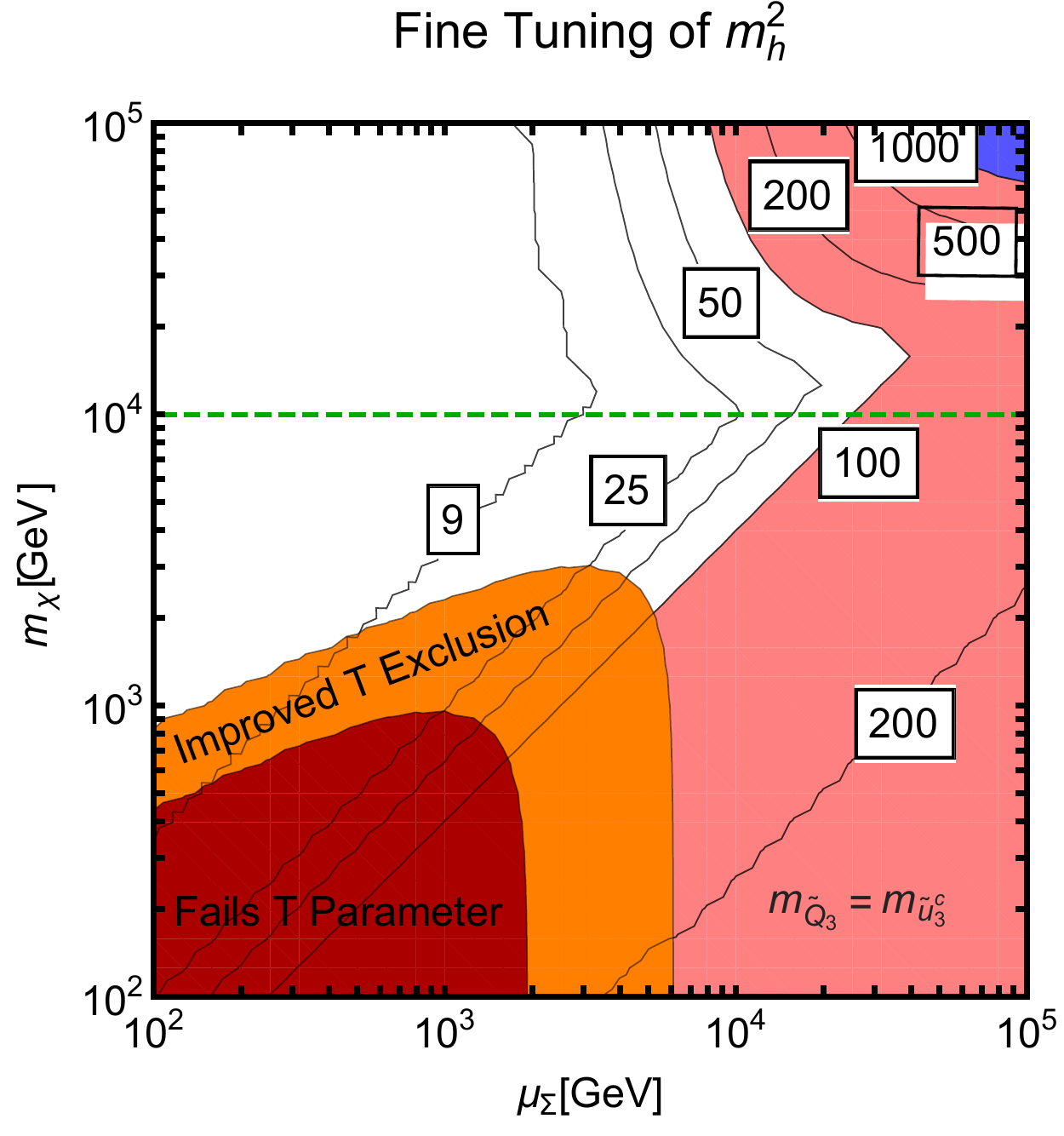}
\caption{Left and right-handed stop masses are equal.}
\label{fig:finetuningTB10BothEqual}
\end{subfigure}
\begin{subfigure}[t]{\textwidth}
\includegraphics[width=0.45\linewidth]{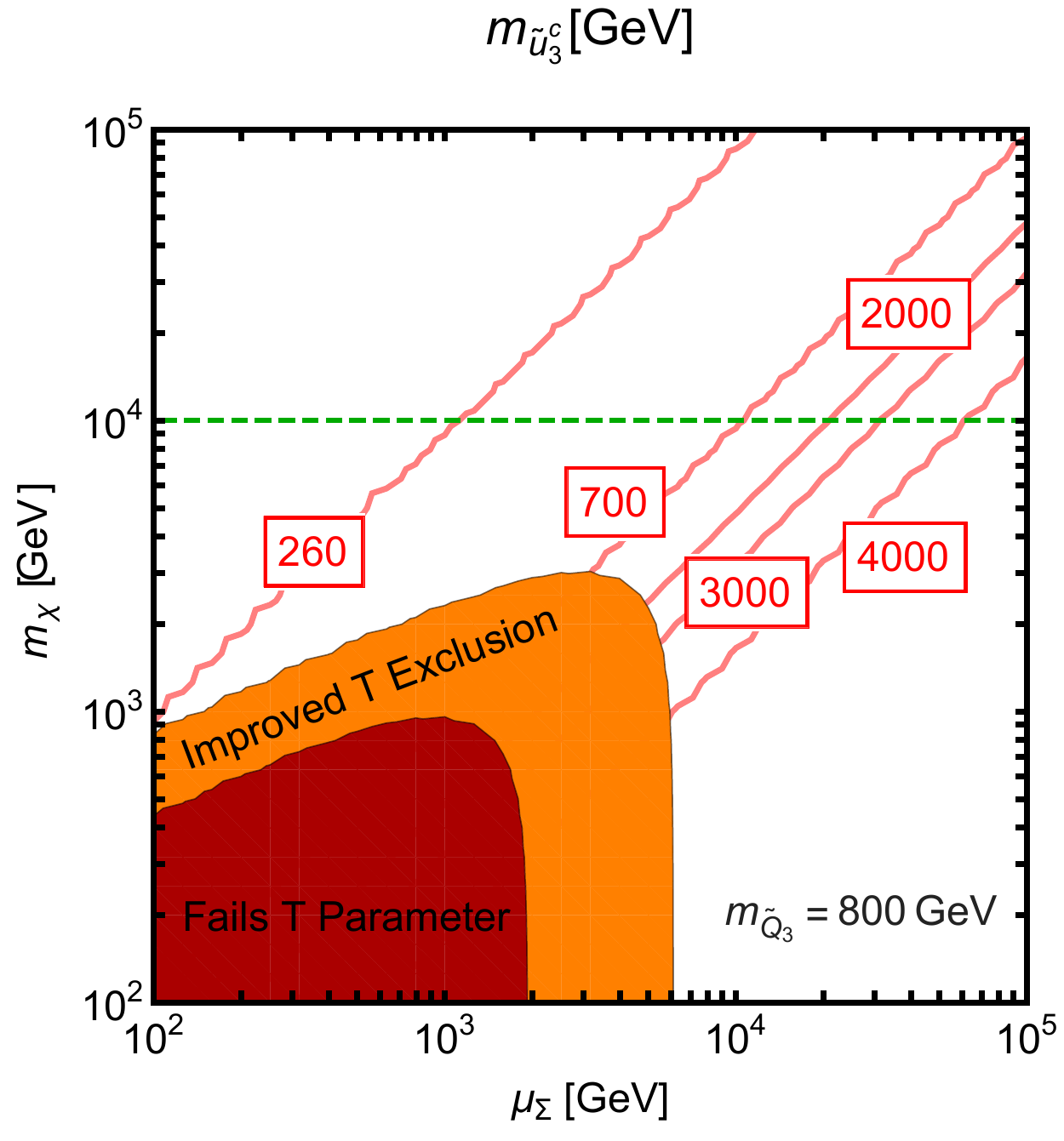}
\includegraphics[width=0.45\linewidth]{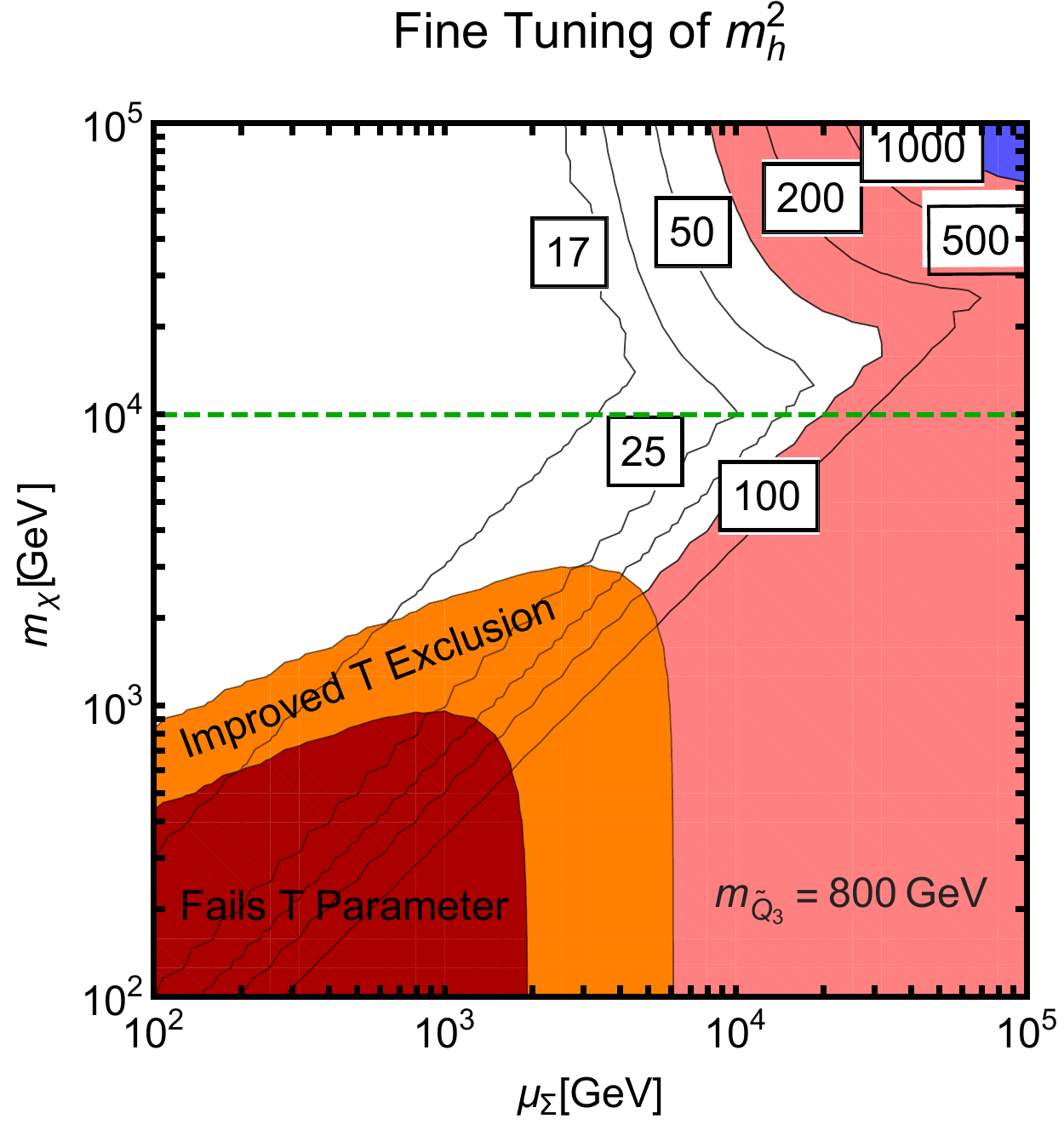}
\caption{Left-handed stop set to $800\gev$ and the right-handed stop sets the Higgs mass.}
\label{fig:finetuningTB10ChangeRight}
\end{subfigure}

\caption{Analogous panels to Fig.\ref{fig:ftLambda}, this time with varying $\mu_{\Sigma}$ and $m_{\chi}$ for fixed $\lambda=0.25$ and $m_{T}=800\gev$. In section \ref{sec:TripPheno}, we study the phenomenology of the dashed green line.}

\label{fig:ft}
\end{center}
\end{figure}

Before discussing the differences between the two different stop assumptions, we scan over $\mu_{\Sigma}$ and $m_{\chi}$ to understand how these affect the Higgs mass, fine tuning, and $\mathcal{T}$. We chose the point
\begin{equation}
\lambda=0.25 ~~~\text{ and }~~~ m_T = 800~\gev,
\end{equation}
which in the first scan lies close to the smallest fine tuned contour and is beyond the reach of the improved $\mathcal{T}$ exclusion. Figure \ref{fig:ft} shows the results of the second scan again with the stop masses in the left panels and the shaded regions the same as in Fig.~\ref{fig:ftLambda}. The triplet contribution to $m_h^2$ is proportional to $m_{\chi}^2/(\mu_{\Sigma}^2+m_{\chi}^2)$. Larger values of $m_{\chi}$ decrease the stop masses while larger $\mu_{\Sigma}$ decouples the effect of the triplets and forces larger stop masses. Lines of constant stop mass run along the diagonal.

The right panels of Fig.~\ref{fig:ft} show the corresponding fine tuning measure. Over most of the parameter space, the fine tuning contours follow the stop mass contours which implies that the RGE running term is dominating the fine tuning. This is not the case in the upper right part of the plots for large values of $m_{\chi}$ and $\mu_{\Sigma}$. In these regions the finite threshold correction piece of the fine tuning dominates. This term is never dominant for $\mu_{\Sigma} \lesssim 1~\tev$ or $m_{\chi} \lesssim 10\tev$. 

The $\mathcal{T}$ parameter constrains more of the parameter space in this scan. In this case, $m_T$ is large so $\bra T^0 \ket$ does not contribute much to $\mathcal{T}$. Instead, $\mathcal{T}$ is controlled by $\bra \chi^0 \ket$ which is proportional to $\mu_{\Sigma}/(\mu_{\Sigma}^2 + m^2_{\chi})$. Keeping the triplet contributions to $\mathcal{T}$ within the 1-$\sigma$ uncertainty excludes out to $\mu_{\Sigma} \le 1.1\tev$ for $m_{\chi} \lesssim 800\gev$. The orange region again shows what could be excluded if the uncertainty were improved by an order of magnitude. This may be the best method for explicitly excluding parameter space and reaches out to $\mu_{\Sigma} \le 1.5\tev$ for $m_{\chi} \lesssim1.2\tev$. Having a low value for $\mu_{\Sigma}$ allows for a large triplet contribution to the Higgs mass without the need to worry about the finite threshold correction term in the fine tuning. In this region, the $\mathcal{T}$ parameter forces $m_{\chi}$ to large values to decrease $\bra \chi^0 \ket$. This in turn \emph{increases} the triplet contribution to the Higgs mass, lowering the fine tuning.

Having discussed how the fine tuning depends on the triplet parameters, we now examine the effects of the different stop assumptions. The general results apply to both scans, but we focus only on the second scan, with $\lambda$ and $m_T$ fixed. The stop contribution to the Higgs mass depends on the geometric mean of the stop masses.  At $\mu_{\Sigma}= m_{\chi} = 10\tev$, the geometric mean of the stops needs to be around $800\gev$. In this case, both assumptions for choosing the stop mass give $m_{\tilde{Q}_3}=m_{\tilde{u}_3^c}=800\gev$ and the corresponding measure of fine tuning is around 50. Lowering the value of $\mu_{\Sigma}$ increases the triplet contributions to the Higgs mass and decreases the stop masses and fine tuning. The minimum stop mass (still along $m_{\chi}=10\tev$) is reached when $\mu_{\Sigma} \le 2 \tev$. When both stop soft masses are simultaneously changed, they take on a minimum mass of around $450\gev$. The minimum fine tuning is then $\Delta\sim9$. On the other hand, when only changing the right-handed soft mass, it needs to be even lighter. Its minimum soft mass is around 260 GeV which gives a fine tuning of 17. Although one stop mass is lighter, the RGE running (and thus the tuning) are worse because the left-handed mass is still at $800\gev$. We have marked the line $m_{\chi}=10\tev$ with a green dashed line and will study the phenomenology along this line in more detail in the next section.

The benchmark values that we have used allow for quite low values of fine tuning for both assumptions about the stop masses. This low fine tuning comes at the cost of having light stops. In fact, for stop mass assumptions, the minimum stop mass achieved is well below the 750-800 GeV LHC limits \cite{Aad:2012xqa,Aad:2014qaa,Aad:2012uu,Aad:2014nra,Aad:2014bva, Aad:2012ywa, Chatrchyan:2012lia, Chatrchyan:2013xna, Chatrchyan:2014lfa, CMS:2014yma, CMS:2014wsa}. In the next section we will show that these searches do not exclude all of our regions of low fine tuning. However, it does raise the question about how the model can deal with LHC SUSY searches and what other signatures to search for. Although the triplet scalars need to be heavy, their fermion counterparts -- the {\em tripletinos} -- with mass $\sim \mu_{\Sigma}$, can be light enough to be reachable by the LHC. In the next section we briefly explore the phenomenology of the tripletinos at the LHC. We will examine both the direct constraints on these particles and how tripletinos affect the decay of the stops.

%*********************Section***********************
\section{Triplet Fermion Phenomenology}
\label{sec:TripPheno}
%*********************Section***********************

%*********************Section***********************
\subsection{(Lack of) Constraints on Tripletinos}
\label{sec:TripConstraints}
%*********************Section***********************

The $Y=1$ triplets contain neutral, $\pm1$ and $\pm2$ charged fermions.  The neutral and singly-charged fermions mix with the neutralinos and charginos, respectively (the mass matrices of the fermions are shown in Appendix \ref{sec:AppMixing}). The doubly-charged states, on the other hand, do not mix with SM particles. One might expect that strong bounds would exist for such exotic states. The tripletinos, however, are good at hiding.

\begin{enumerate}
\item Direct Searches

	The charge $\pm 1, 0$ tripletinos are subject to MSSM electroweakino searches, which currently exclude regions where the LSP mass is less than around 150 GeV if there are no light sleptons \cite{Aad:2014vma,CMS:2013dea}. These searches are most powerful if the LSP is light and if there is a large separation between the mass of the LSP and the mass of the rest of the other states. As a result, these conventional searches fail for quasi-degenerate electroweakino spectra, such as one expects in a pure Higgsino scenario or with a Higgsino-tripletino admixture. Another possibility is to look for disappearing tracks \cite{CMS:2014gxa} or long-lived charged particles \cite{Aad:2013pqd,Chatrchyan:2013oca}, though these approaches require a level of degeneracy that is atypical in the region of tripletino-Higgsino parameter space we are interested in.
	
One potential avenue is a search focusing on the doubly charged tripletinos and $\mu_{\Sigma} < \mu$.  The (lighter) mass eigenstates are then given by
	\begin{equation}
	\begin{aligned}
		m_{\tilde{\chi}^{++}} &=\mu_{\Sigma}, \\
		m_{\tilde{\chi}^{+}} &= \mu_{\Sigma} \left(1-\frac{1}{2}\frac{\lambda^2 v^2}{\mu^2} (1-\cos(2\beta) \right), ~~\text{and} \\
		m_{\tilde{\chi}^0} &= \mu_{\Sigma} \left(1-\frac{\lambda^2 v^2}{\mu^2} (1-\cos(2\beta) \right).
	\end{aligned}
	\end{equation}
	For  benchmark parameters $\mu=250\gev, \lambda = 0.25$ and taking $\mu_{\Sigma}=150~\gev$, the masses are 150, 145.5, and 141 GeV respectively. The pair production cross section of the doubly charged state at the LHC is $1.05$ ($2.48$) pb for the LHC at $8$ ($14$) TeV. These decay down to the neutral state through $W^{\pm}$ bosons. Although the decay products will be soft and hard to detect, the signal has 4 $W^{\pm}$ bosons which can decay leptonically. A dedicated search is beyond the scope of this paper, but the relatively large cross section along with the clean final state could motivate a search for the doubly charged particles -- recoiling off a hard, initial-state jet for triggering purposes.
	
\item Oblique parameters

	Triplet fermions have the potential to generate a loop level contribution to the $\mathcal T$ parameter. However, at $O(\lambda^2)$ we find this contribution to be zero due to the Dirac nature of the tripletinos and the near degeneracy of the states. We calculated this using mass insertions to account Higgsinos-tripletino mixing, as well as in an effective theory where the Higgsinos were integrated out. In both cases the vacuum polarization amplitudes $\Pi^{11}(0)$ and $\Pi^{33}(0)$ are non-zero, but their difference is zero.

\item Higgs observables

	 The addition of $SU(2)_L$ triplets to the content of the MSSM adds more charged particles which couple to the Higgs and could affect the decay of $h\rightarrow \gamma\gamma$. Unlike more traditional triplet extensions \cite{Delgado:2012sm,Delgado:2013zfa, Kang:2013wm, Arina:2014xya} only one of the triplets couples to Higgses, and in the $Y=\pm1$ Dirac Triplet extension of the MSSM, the partial width is not affected to lowest order. The only way that the triplets in this model play a role in the diphoton rate is allowing for lower stop masses which affect both the production and the decay of the Higgs \cite{Carena:2011aa, Carena:2013iba}.

\end{enumerate}

Moving to direct production at the LHC, the triplet fermions are hard to detect due to the small mass splitting. Giving the triplets a Dirac mass and having only one triplet couple to the doublet makes their presence hard to find in sensitive loop level processes too. The effects of the triplets can still be seen in the efficient raising of the Higgs mass leading to light stops. If the triplet fermions happen to be lighter than the stops it would be possible use stop decays to observe the triplet fermions.

%*********************SubSection***********************
\subsection{Stop Decays}
\label{subsec:StopDecays}
%*********************SubSection***********************

We have seen that the inclusion of $Y = \pm 1$ triplets with interactions inspired by the DiracNMSSM -- namely where only one triplet couples to Higgses -- leads to light stops. While nice from a fine-tuning perspective, light stops are constrained by the LHC, so we must make sure these `natural' scenarios are not ruled out by experimental searches. As we illustrate in this section, the phenomenology of the stops depends on the hierarchy of $\mu$ and $\mu_{\Sigma}$ and whether the lightest stop is left or right-handed. In all four scenarios we sketch out the viable parameter space. In most circumstances, we find that compressed spectra are required to avoid LHC limits, such that larger values of $\mu$ are necessary; this a posteriori motivates our benchmark choice $\mu = 250\,\gev$. 

To anchor our phenomenology study, we fix $\lambda=0.25$, $m_T=800\gev$ $m_{\chi} = 10\tev$, and vary $\mu_{\Sigma}$ (all other parameters are taken from Table~\ref{tab_BenchmarkFT}). This parameter slice is indicated by the green dashed line in Figs.~\ref{fig:finetuningTB10BothEqual} and \ref{fig:finetuningTB10ChangeRight} and is characterized by low fine-tuning. The spectrum of the charginos, neutralinos and stops along this line is shown below in Fig.~\ref{fig:Spectrum}. The solid colored lines show the chargino/neutralino masses; the sharp feature at $\mu_{\Sigma} \sim \mu = 250\,\gev$ corresponds to where the composition of the lightest $\tilde{\chi}^0_i, \tilde{\chi}^+$ shifts from primarily tripletino to primarily Higgsino.
\begin{figure}[h!]
\begin{center}
\includegraphics[width=0.45\linewidth]{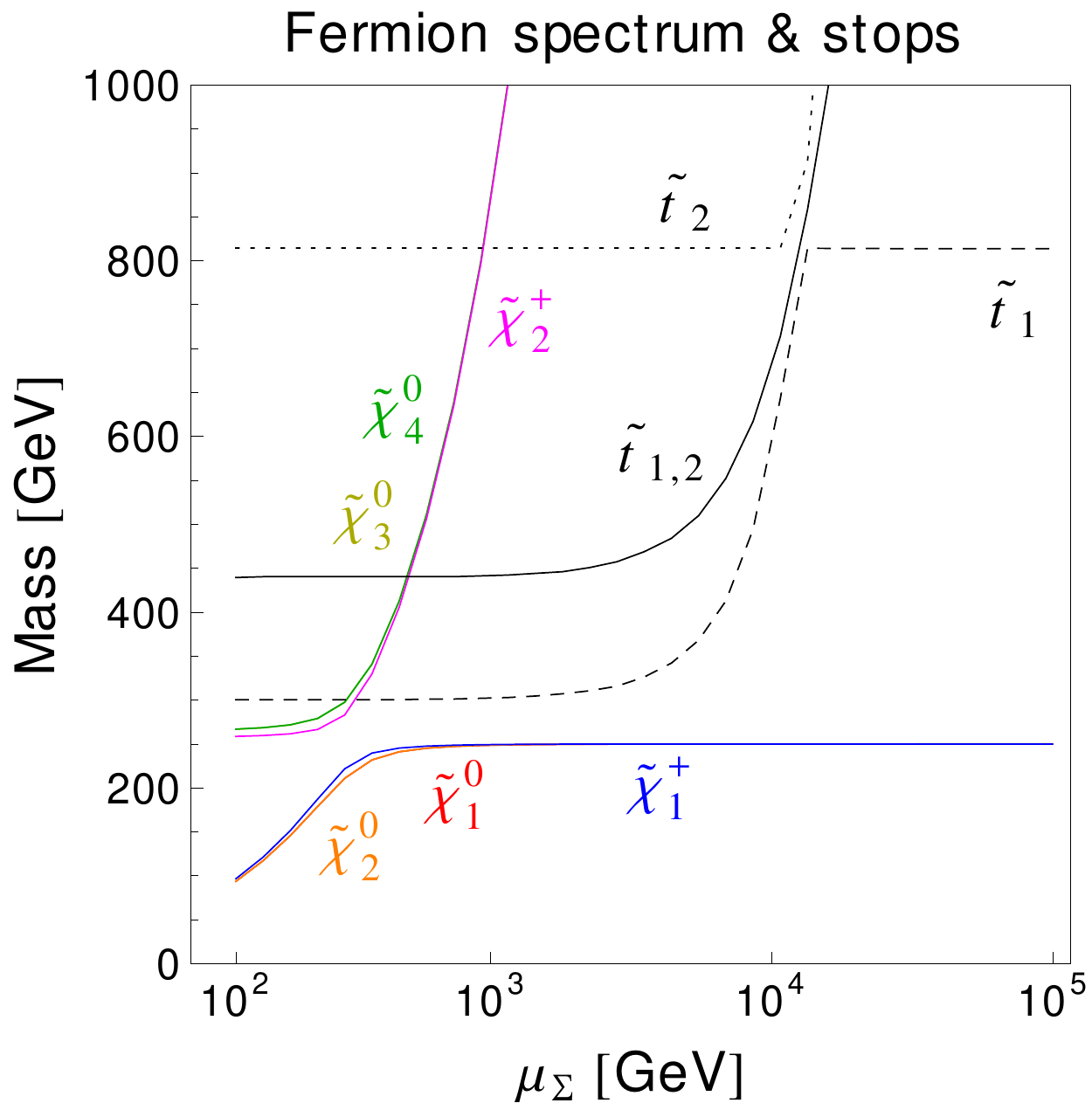}
\caption{Spectrum of the stops, neutralinos and charginos. The Higgsino mass parameter $\mu=250\gev$ while the triplet mass is along the horizontal axis. Two methods of choosing the stop mass are shown. The solid black line labelled $\tilde{t}_{1,2}$ marks changing both the left and right soft masses simultaneously. The dashed lines keep the left-handed soft mass at $800\gev$ and use the right-handed mass to set the Higgs mass.}
\label{fig:Spectrum}
\end{center}
\end{figure}
The black lines in Figure~\ref{fig:Spectrum} indicate the stop spectra for both stop selection choices (see Sec.~{\ref{sec:Num}}). The solid line corresponds to changing both the left and the right-handed soft masses simultaneously. The dashed line, labeled $\tilde{t}_1$, and the dotted line, labelled $\tilde{t}_2$ mark the masses of the two stops when the left-handed soft mass is set to $800\,\gev$ and the right-handed mass moves to accommodate the Higgs mass. 

The next ingredient in the stop phenomenology is the branching ratio. Using the same set of parameters as in Fig.~\ref{fig:Spectrum}, we plot the branching ratio below in Fig.~\ref{fig:BranchingRatio} for both stop scenarios. In the branching ratio calculations we only keep the two-body final states. 
\begin{figure}[h!]
\begin{center}
\includegraphics[width=0.45\linewidth]{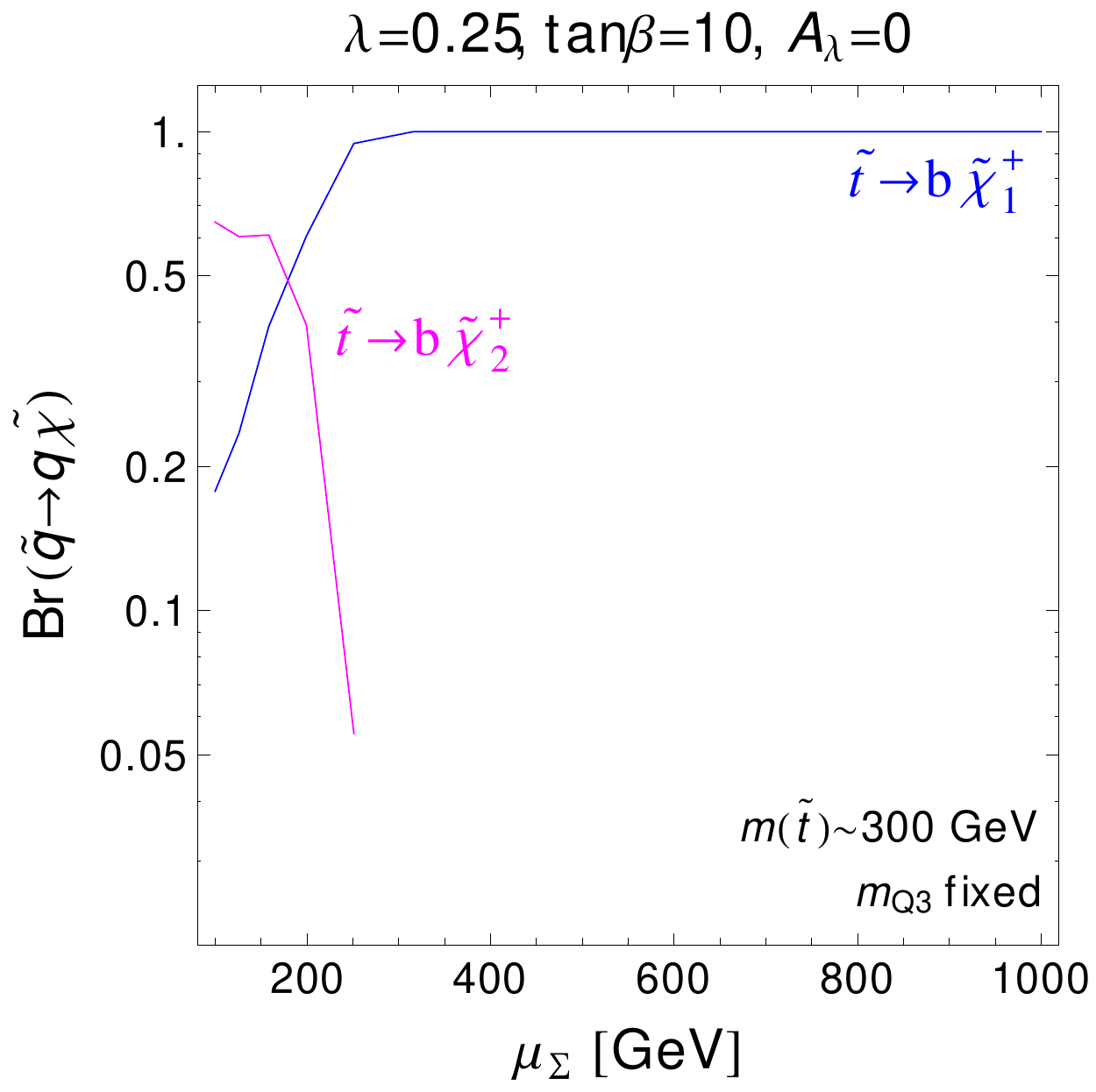}
\includegraphics[width=0.45\linewidth]{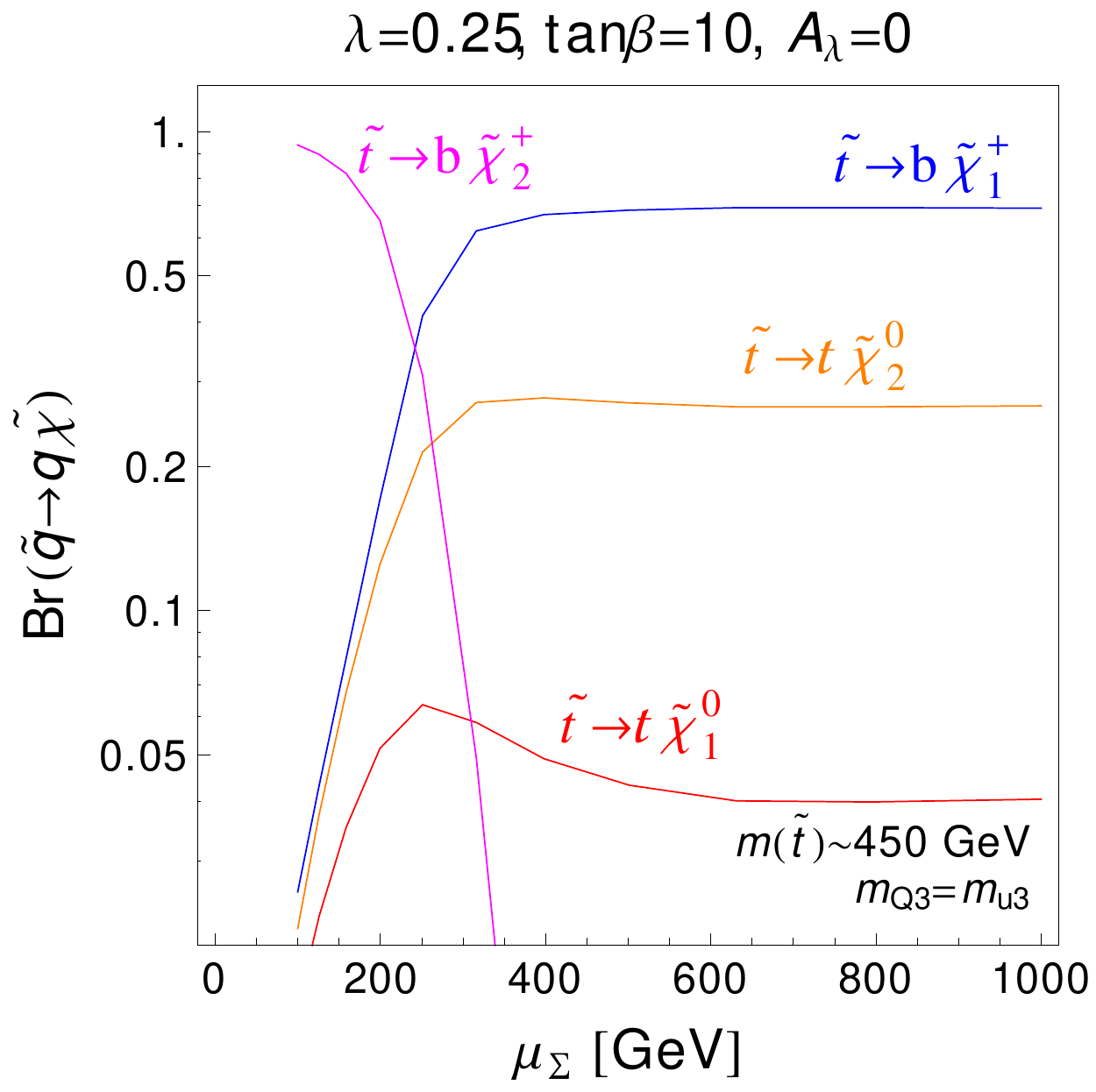}
\caption{Branching ratios of the stops when only considering 2-body decays. The bino and wino have been completely decoupled, only leaving the Higgsino and tripletino for the stop decays. The left panel has the left-handed stop mass set to $800\gev$ and uses the right-handed mass to raise the Higgs mass. The right panel has both soft masses change to set the Higgs mass. The Higgsino mass is $\mu=250\gev$.
}
\label{fig:BranchingRatio}
\end{center}
\end{figure}
Both sets of branching ratios show a feature at $\mu_{\Sigma} \sim 250\,\gev$ where the character of the electroweakinos changes. For light right-handed stops (left panel of \ref{fig:BranchingRatio}) the branching fraction for $\tilde t_1 \to b\,\chi^+_1$ is $\sim100\%$ over a wide range of $\mu_{\Sigma}$ because the triplet states do not couple directly to the stops and the stop mass in this scenario is nearly the same mass as our benchmark Higgsino (the LSP) mass. In the right panel, where both left and right-handed stops have the same mass, there is more variety in the branching ratios because the stops are heavy enough to undergo both $\tilde t \to t\, \chi^0_i$ and $\tilde t \to b\, \chi^+_i$ decays. 

From Figs.~\ref{fig:Spectrum} and \ref{fig:BranchingRatio}, we can see the phenomenology naturally splits up into four categories, $\mu < \mu_{\Sigma}, \mu > \mu_{\Sigma}$ for either $m_{\tilde t_1} \ll m_{\tilde t_2}$ or $m_{\tilde t_1} \cong m_{\tilde t_2}$, which we discuss in more detail below:\\

{\it Case $m_{\tilde t_1} \ll m_{\tilde t_2}$:} Here the left-handed stop mass is fixed to $800\,\gev$ and the right-handed stop mass is variant to satisfy the Higgs mass. For $\mu_{\Sigma} \lesssim 2\,\tev, m_{\tilde t_1} \sim 300\,\gev$.

\begin{itemize}
\item $\mu_{\Sigma} > \mu$: Here the tripletinos play little role, and the low energy states are simply stops and Higgsinos. These scenarios are tightly constrained unless the Higgsino mass $\mu$ is nearly the same as the stop mass and the only two-body decay mode is $\tilde t \to b\,\chi^{+}_1$.  As $\mu$ approaches $m_{\tilde t_1}$, the $b$ and subsequent $\chi^+_1$ decay products become soft and conventional stop searches become inefficient. For $m_{\tilde t_1} = 300\,\gev$, a Higgsino mass of $\mu \gtrsim 180\,\gev$ is needed~\cite{Aad:2014nra, CMS:2014yma, CMS:2014wsa, Kribs:2013lua} to avoid current LHC bounds.

\item $\mu_{\Sigma} < \mu$: In this case the tripletinos are lighter than the Higgsinos, so stop decays proceed in two steps; stop decaying to Higgsino, then Higgsino decaying to tripletino. The visibility of this setup depends on the $\mu_{\Sigma} - \mu$ difference. If the two scales are sufficiently separated, the Higgsino decays are energetic and will be picked up by standard stop searches, regardless of how degenerate $\mu$ and $m_{\tilde t_1}$. Therefore, for this scenario to be viable, all three scales $m_{\tilde t_1}, \mu$ and $\mu_{\Sigma}$ must be nearby; for the benchmark value $\mu = 250\,\gev$, we estimate $\mu_{\Sigma} \gtrsim 200\,\gev$ is required.
		
\end{itemize}

{\it Case $m_{\tilde t_1} \sim m_{\tilde t_2}$:} In this case, the stop masses are changed together to accommodate the Higgs mass. The stops have a mass of around $450\gev$ for $\mu_{\Sigma}\lesssim 2\,\tev$. For larger $\mu_{\Sigma}$,  the triplet contribution to $m^2_h$ shrinks and the stops quickly increase in mass.

\begin{itemize}
\item $\mu_{\Sigma} > \mu$: The stop now has phase space to decay through a top quark and does so around 30$\%$ of the time. Searches for this mode include the leptonic decays and all hadronic decays \cite{Aad:2014bva, Aad:2012ywa, Chatrchyan:2012lia, Chatrchyan:2014lfa}. For a stop mass of 450 GeV, the limits extend up to an LSP mass of around 220 GeV, thus our model with $\mu=250\gev$ survives. However, a left-handed stop implies a left-handed sbottom of similar mass. The sbottom searches are very effective for this sort of the spectrum and place constraints on the sbottom up to a mass of $\sim700\gev$ for an LSP mass of $250\gev$~\cite{ Aad:2013ija, CMS:2014nia}. The sbottom (and stop) mass is raised above $700\gev$ when the triplet effects are decoupled with $\mu_{\Sigma} > 10\tev$. In the large region of parameter space where the sbottoms are $450\gev$, in order to be viable, the LSP mass ($\mu$ in this case) must be raised to $\sim300\gev$.

\item $\mu_{\Sigma} < \mu$: In this region, all stops and bottoms first decay to Higgsino plus $b/t$, with the Higgsino subsequently decaying to tripletino. The sbottom searches can again be useful, but one potential caveat is that the sbottom decays in our scenario are quite busy, containing extra objects from the Higgsino decay. These final states may be inefficient in sbottom searches such as \cite{CMS:2014nia} which explicitly veto events with leptons or with more than two jets. The extent to which this scenario can evade the sbottom searches without being collected by another search requires a dedicated analysis, though it is possible that a region window near $\mu_{\Sigma} \sim\mu$ exists undetected by current stop or sbottom searches. 

\end{itemize}

Summarizing, the light stops that are a consequence of this triplet extension are safe from current LHC bounds if the spectrum is sufficiently squeezed. For $m_{\tilde t_1} \ll m_{\tilde t_2}$ (light right-handed stop), the benchmark ($\mu = 250\gev$) scenario is safe provided $\mu_{\Sigma} > 200\gev$. For degenerate left and right-handed stops, the bounds are more stringent and are driven by sbottom searches. For the benchmark set of parameters to be safe, either the entire stop spectrum must be raised to $\gtrsim 700\gev (\mu_{\Sigma} > 10\tev)$ or the Higgsinos and tripletinos must be made more degenerate with the stops, $\mu_{\Sigma} \sim \mu \gtrsim 300\gev$. Continued searches for stop and sbottom squarks will place tighter constraints on the model if no sparticle is found. These stop limits may be alleviated, for example by lowering $\lambda$ or raising $\mu$, though at the expense of increased fine tuning.

%*********************Section***********************
\section{Discussion and conclusion}
\label{sec:conclusions}
 %*********************Section***********************
 
 We have examined extensions of the MSSM by two $SU(2)_L$ triplets where only one triplet is permitted to couple to the Higgs doublets. While not generic, this setup is radiatively stable and has the property -- first pointed out in the DiracNMSSM~\cite{Lu:2013cta} using singlets -- that large, $\gtrsim\,\text{few}\,\tev$ soft masses for the uncoupled field generate tree level contributions to the Higgs mass without the price of increased fine tuning. Triplet extensions can either have $Y = 0$ or $Y = \pm 1$, we have a studied the Higgs mass contributions, fine tuning, and $\mathcal T$-parameter constraints for both cases. 
 
 Triplets with nonzero hypercharge are well-suited to this scenario as they must appear in pairs and can only have Dirac-type superpotential masses. For $Y = \pm 1$ scenarios, we find $m_h = 125\,\gev$ can be achieved with fine tuning as small as one part in ten (according to the same fine tuning measure used in~\cite{Lu:2013cta}). We find that the least tuned regions of parameter space coincide with regions where the $\mathcal T$-parameter constraint -- usually a thorn in the side of triplet models -- is not an issue. The smallness of the $\mathcal T$-parameter is a consequence the $\tan\beta$ dependence of the triplet-Higgs interaction, aided by the fact that the uncoupled triplet soft mass can be very large ($\gtrsim \tev$).
 
The least tuned regions also have light stop spectra, either  $m_{\tilde t_1} \sim300\,\gev$ or  $m_{\tilde t_1} \sim 450\,\gev$ depending on whether only one stop is light or both. Such light stops are running out of hiding places at the the LHC. In order to remain undetected, the stops must be fairly degenerate with the LSP, $m_{\tilde t_1} - m_{LSP} \lesssim 100\,\gev$, though the details of the bounds depend on the hierarchy of the triplet Dirac mass $\mu_{\Sigma}$ and the Higgsino mass $\mu$, as well as on the handedness of the lightest stop; scenarios with light right-handed stops are less constrained than with left-handed.  

In addition to light stops, the charged and neutral fermionic components of the triplets, the tripletinos, may be light. In the parameter space of interest for the purposes of raising the Higgs mass, these triplets are unconstrained by existing LHC searches. This stealthiness is due to the small splitting among the triplet states and because the tripletinos only couple to Higgs and gauge bosons at tree level. Finally, for certain triplet parameters -- for example $\mu_{\Sigma} \sim m_{\chi} \sim 2\,\tev$ for the parameter set in Fig.~\ref{fig:finetuningTB10ChangeRight}, the $\mathcal T$-parameter contribution from the triplet sector may be within the reach of future precision electroweak studies.

\subsection*{Acknowledgments}
The work of AD was partially supported by the National Science Foundation under Grant No. PHY-1215979, and the work of AM was partially supported by the National Science Foundation under Grant No. PHY-1417118.

\appendix

\section{Potential for the $Y=0$ triplets}
\label{sec:appY0}

In the following two appendices we list the effective potential, the expressions for the soft masses in terms of the model parameters (via the minimization conditions) and the change in the Higgs mass coming from the triplet sector. It must be emphasized that all of these are tree-level quantities that will receive loop corrections. For the model involving two $Y=0$ triplets, the triplet fields are given by
\begin{equation}
\begin{aligned}
\Sigma_{1} &= \begin{pmatrix} T^{0}/\sqrt{2} & -T_{2}^{+} \\ T_{1}^{-} & -T^{0}/\sqrt{2} \end{pmatrix}  \text{ and}\\
\Sigma_{2} & = \begin{pmatrix} \chi^0/\sqrt{2} & -\chi^+_2 \\ \chi^-_1 & -\chi^0/\sqrt{2} \end{pmatrix}.
\end{aligned}	
\end{equation}
The only change in the superpotential from the MSSM is
\begin{equation*}
W \supset \lambda H_u \cdot \Sigma_2 H_d.
\end{equation*}
Expanding the neutral scalar potential including the soft terms leads to
\begin{align}
V _{\text{neutral}}
&= m^2_{H_u}|H_u^0 |^2 + m^2_{H_d^0}|H_d^0|^2 + m^2_{\chi}|\chi^0|^2 + m^2_{T}|T^0|^2 \notag \\
& +\left| \frac{\lambda}{\sqrt{2}} H_d^0 T^0 - \mu H_d^0 \right|^2 + \left| \frac{\lambda}{\sqrt{2}} H_u^0 T^0 - \mu H_u^0 \right|^2 \notag \\
&+ \left| \mu_{\Sigma} \chi^0 + \frac{\lambda}{\sqrt{2}} H_d^0 H_u^0 \right|^2 + \left|\mu_{\Sigma} T^0 \right|^2 + \frac{g^2 + g^{\prime 2}}{8} (|H^0_d|^2 - |H^0_u|^2)^{2} \notag \\
&+ \left(\mu_{\Sigma} B_{\Sigma} \chi^0T^0 + B_{\mu} \mu  H_d^0 H_u^0 + \frac{A_{\lambda}\lambda}{\sqrt{2}} H_d^0 H_u^0 \chi^0 + \text{h.c.} \right). \label{eqn:NeutralPotentialY0}
\end{align}

The heavy triplet scalars are then integrated out, leading to an effective potential of
\begin{align}
V_{\text{eff}}
&\supset \left(m^2_{H_u} + |\mu|^2\right)|H_u^{0}|^2 + \left(m^2_{H_d} + |\mu|^2\right)|H_d^{0}|^2 \notag \\
&+  \frac{m_Z^2}{4 v^2} (|H_d^0|^{2}-|H_u^0|^{2})^2 - \left(B_{\mu} \mu H_d^0 H_u^0 + \text{ h.c.}\right) \notag \\
&+ \frac{\left| \lambda H_d^0 H_u^0 \right|^2}{2} \left(1- \frac{\mu_{\Sigma}^2}{\mu_{\Sigma}^2 + m^2_{\chi}} \right) \notag \\
&- \frac{\lambda^2}{2 (\mu_{\Sigma}^2 + m^2_{T})} \left|A_{\lambda} H_d^0 H_u^0 -\mu \left( |H_u^0|^2 + |H_d^0|^2 \right) \right|^2  + (\text{higher order})  
\label{eqn:VintoutY0} .
\end{align}
Terms of order $O(D_{\chi}^{-2},D_{T}^{-2},D_{\chi}^{-1}D_{T}^{-1})$ and higher inverse powers have been neglected, where $D_{\chi,T}\equiv (\mu_{\Sigma}^{2}+m_{\chi,T}^{2})$.
The conditions needed to achieve EWSB at the minimum of this potential are
\begin{eqnarray}
m^2_{H_u} &=& -|\mu|^2 + \frac{m^2_Z}{2} \cos(2\beta) + m^2_A \cos^2 \beta -\frac{\lambda^2v^2}{2} \cos^2 \beta \nn 
	& &+ \frac{v^2 \lambda^2}{2} \frac{- 4 |\mu|^2 - A_{\lambda}(\mu + \mu^*)(\cos(2\beta)-2)\cot\beta - 2 A_{\lambda}^2 \cos^2 \beta}{\mu_{\Sigma}^2 + m^2_{T}} \nn 
	&& - \mu_{\Sigma}^2 v^2 \lambda^2 \frac{\cos^2 \beta}{\mu_{\Sigma}^2 + m^2_\chi} \text{, and} \label{eqn:mincondHu} \\
m^2_{H_d} &=& -|\mu|^2 -\frac{m^2_Z}{2} \cos(2\beta) + m^2_A \sin^2 \beta - \frac{\lambda^2 v^2}{2} \sin^2\beta \nn 
 && + \frac{v^2 \lambda^2}{2} \frac{-4 |\mu|^2 + A_{\lambda} (\mu + \mu^*)(2+\cos(2\beta))\tan\beta - A_{\lambda}^2 \sin^2 \beta} {\mu_{\Sigma}^2 + m^2_{T}} \nn
 && - \mu_{\Sigma}^2 v^2 \lambda^2 \frac{\sin^2\beta}{\mu_{\Sigma}^2 + m^2_{\chi}} \label{eqn:mincondHd}
\end{eqnarray} 
The corresponding shift in the MSSM physical Higgs mass in the decoupling limit
\begin{equation*}
\Delta m_{h}^{2}=\frac{v^2 \lambda^2}{2}\sin^2(2\beta) \frac{m^2_{\chi}}{\mu^2_{\Sigma}+m^2_{\chi}} -\frac{v^2\lambda^2}{2} \frac{\left|2 \mu^* -A_{\lambda} \sin(2\beta)\right|^2}{\mu^2_{\Sigma}+m^2_{T}}.
\end{equation*}

\section{Potential for the  $Y=\pm1$ triplets}
\label{sec:appY1}

Now we examine the model where the triplets have hypercharge $Y=\pm1$, which can then be expressed as
\begin{equation}
\begin{aligned}
\Sigma_{1}
&= \left(\begin{array}{cc}
T^{-}/\sqrt{2} & -T^{0} \\
T^{--} & -T^{-}/\sqrt{2}
\end{array}\right) \text{ and}
\\
\Sigma_{2}&= \left(\begin{array}{cc}
\chi^{+}/\sqrt{2} & -\chi^{++} \\
\chi^{0} & -\chi^{+}/\sqrt{2} \end{array} \right).
\end{aligned}
\end{equation}
The superpotential is modified from the MSSM with
\begin{equation*}
W \supset \lambda H_u \cdot \Sigma_1 H_u
\end{equation*}
The neutral potential is then given by
\begin{eqnarray}
V_{\text{neutral}} &= & m^2_{H_u} \left| H_u \right|^2 + m^2_{H_d} \left| H_d \right|^2 + m_{\chi}^2  {|\chi^0|^2} + m_{T}^2  {|T^0|^2} \nn
	&&+ \left|2 \lambda H_u^0 T^0 + \mu H_d^0 \right|^2 +\left|\mu H_u^0\right|^2 +\left|\mu_{\Sigma} T^0\right|^2 + \left| \mu_{\Sigma} \chi^0 + \lambda H_u^0 H_u^0 \right|^2 \nn
	&&+ \frac{g^2 + g^{\prime~2}}{8} \left( H_d^0 H_d^{0*} - H_u^0 H_u^{0*} + 2 T^0 T^{0*} - 2\chi^0 \chi^{0*} \right)^2 \nn
	& &+ \left( - \lambda A_{\lambda} H_u^0 H_u^0 T^0 - \mu B_{\mu} H_d^0 H_u^0 - \mu_{\Sigma} B_{\Sigma} T^0 \chi^0 + \text{h.c.} \right)
\end{eqnarray}
The heavy triplets are integrated out, leaving an effective potential of
\begin{eqnarray}
	V_{\text{eff,neut}} &=& \left(m^2_{H_u} + \mu^2 \right) |H_u^0|^2 + \left(m^2_{H_d} + \mu^2 \right) |H_d^0|^2 \nn 
	&&+ \frac{m_Z^2}{4 v^2} \left( |H_d^0|^2 - |H_u^0|^2 \right)^2 -\left( \mu B_{\mu} H_d^0 H_u^0 +\text{h.c.}\right) \nn
	&&+ \lambda^2 |H_u^0H_u^0|^2 \left(1 - \frac{2A_{\lambda}^2}{\mu_{\Sigma}^2 +m^2_{T}} - \frac{2\mu_{\Sigma}^2}{\mu_{\Sigma}^2 +m^2_{\chi}} \right)\nn
	&& -8 |H_u^0|^2 |H_d^0|^2 \lambda^2 \mu^2 \frac{1}{\mu_{\Sigma}^2+m^2_{T}}\nn
	&& + \frac{4 \lambda^2 A_{\lambda}}{\mu^2_{\Sigma} + m^2_{T}} \left(\mu^* H_u^0 H_u^{0*} H_u^{0*} H_d^{0*} + \text{h.c.} \right)  + \mathcal{O}(\frac{1}{D_\chi^2},\frac{1}{D_\chi D_T}, \frac{1}{D_T^2}).
\label{eqn:VintoutY1}
\end{eqnarray}
The minimization conditions are given by
\begin{eqnarray}
m^2_{H_u} &=&- |\mu|^2 +\frac{1}{2}m^2_Z \cos(2\beta) +m^2_A \cos^2(\beta) - 2 v^2 \lambda^2 \sin^2(\beta) \nonumber \\
 &&+ 2 \sin^2(\beta) \frac{\mu^2_{\Sigma} v^2 \lambda^2 }{\mu^2_{\Sigma} + m^2_{\chi}} +2 v^2\lambda^2 \sin^2(\beta) \frac{A_{\lambda}^2 + 2 \mu^2 \cot^2(\beta) + 2 A_{\lambda} \mu \cot(\beta) }{\mu^2_{\Sigma}+m^2_{T}} \label{eqn:mincondHuY1}\\
m^2_{H_d} &=& -|\mu|^2 -\frac{1}{2}m^2_Z \cos(2\beta) - m^2_A \sin^2(\beta) + 4 \frac{|\mu|^2 v^2 \lambda^2 \sin^2(\beta)}{\mu^2_{\Sigma}+m^2_{T}}.
\label{eqn:mincondHdY1}
\end{eqnarray}
This leads to a shift in the Higgs mass in the decoupling limit of
\begin{equation}
\Delta m_{h}^{2}=4 v^2 \lambda^2 \sin^4(\beta)\left( \dfrac{ m_{T_1}^2}{\mu_{\Sigma}^2+m^2_{\chi}} \right)  -\dfrac{4 v^2 \lambda^2 \sin^2{(\beta)}}{\mu^2_{\Sigma} +m^2_{T}}\left|2\mu^* \cos {(\beta)} - A_{\lambda} \sin{(\beta)}\right|^2.
\end{equation}

\section{Finite threshold correction}
\label{sec:AppFTC}

The threshold correction arises when the heavy triplet fields are integrated out. The one loop contribution is given by
\begin{align}
\delta m_{H_u^0}^2
&= \frac{8(\lambda^2 + \frac{\lambda^2}{2}) \mu_{\Sigma}^2 }{16\pi^2} \left(\frac{2}{\epsilon} - \gamma  + 1 + \log 4\pi - \log(\mu_{\Sigma}^2) \right) \notag \\
& + \left(4 \lambda^2 \mu{\Sigma}^2 + 2 \lambda^2 \mu_{\Sigma}^2 \right)\frac{1}{16\pi^2} \left( -\frac{2}{\epsilon} +\gamma -1 -\log4\pi  + \log(m_{\chi}^2 + \mu_{\Sigma}^2) \right) \notag \\
&+ \lambda^2 (4 + 2)  \left(\mu_{\Sigma}^2+m_{T}^2 \right) \frac{1}{16\pi^2} \left(-\frac{2}{\epsilon} +\gamma -1 -\log4\pi  + \log(m_{T}^2 + \mu_{\Sigma}^2) \right) \notag \\
=& \frac{12 \lambda^2 \mu_{\Sigma}^2}{16 \pi^2} \left(\frac{1}{2} \log (m^2_{\chi} + \mu_{\Sigma}^2) +\frac{1}{2} \log (m^2_{T} + \mu_{\Sigma}^2) - \log(\mu_{\Sigma}^2) \right) \notag \\
&+ \frac{6 \lambda^2 m^2_{T}}{16\pi^2} \left(-\frac{2}{\epsilon} +\gamma -1 -\log4\pi  + \log(m_{\chi}^2 + \mu_{\Sigma}^2) \right) \notag \\
=& \frac{6 \lambda^2 \mu_{\Sigma}^2}{16 \pi^2} \left(\log \frac{(m^2_{\chi} + \mu_{\Sigma}^2)}{\mu_{\Sigma}^2} + \log \frac{(m^2_{T} + \mu_{\Sigma}^2)}{\mu_{\Sigma}^2} \right) + \frac{6 \lambda^2 m^2_{T}}{16\pi^2} \left(-\frac{2}{\epsilon} +\gamma -1 -\log4\pi  + \log(m_{T}^2 + \mu_{\Sigma}^2) \right).
\end{align}
We are only interested in the finite piece.

\section{Neutralino and chargino mixing in $Y=\pm1$}
\label{sec:AppMixing}

The $Y=\pm1$ mixing matrix for the neutralinos in the basis $\psi^0=\left( \widetilde{B},\widetilde{W}^{0},\widetilde{H_{d}^{0}},\widetilde{H_{u}^{0}},\widetilde{T}^{0},\widetilde{\chi}^{0} \right)$ is given by
\begin{eqnarray}
\Lc_{\text{Neutralino Mass}} &=& -\frac{1}{2} (\psi^0)^T \mathbf{M}_{\tilde{N}} \psi^0 + \text{c.c.} \\
 \mathbf{M}_{\tilde{N}} &=&
  \begin{pmatrix}
  M_{1}                &                    0 & -c_{\beta}s_{W}m_{Z} & s_{\beta}s_{W}m_{Z}  &     -\sqrt{2}g'v_{T} & \sqrt{2}g'v_{\chi} \\
                   0 &                M_{2} &  c_{\beta}c_{W}m_{Z} & -s_{\beta}c_{W}m_{Z} &      -\sqrt{2}gv_{T} & \sqrt{2}gv_{\chi} \\
-c_{\beta}s_{W}m_{Z} &  c_{\beta}c_{W}m_{Z} &                    0 &       -\mu &                    0 & 0  \\
 s_{\beta}s_{W}m_{Z} & -s_{\beta}c_{W}m_{Z} &       -\mu &       -2v_{T}\lambda & -2v\lambda s_{\beta} & 0 \\
    -\sqrt{2}g'v_{T} &             -\sqrt{2}gv_{T}  &            0 & -2v\lambda s_{\beta} &                    0 & -\mu_{\Sigma} \\
  \sqrt{2}g'v_{\chi} &           \sqrt{2}gv_{\chi}  &            0 &                    0 &        -\mu_{\Sigma} & 0
  \end{pmatrix}, \nonumber
\label{neutralinoMixY1}
\end{eqnarray}
where $c_{\beta}$, $s_{\beta}$, $c_W$,  and $s_W$ represent the cosine or sine of beta or $\theta_W$. 
The triplets add one chargino. Using the basis $\psi^{\pm} = \left( \widetilde{W}^{+},\widetilde{H}_{u}^{+},\widetilde{\chi}^{+}, \widetilde{W}^-, \widetilde{H}_d^-, \widetilde{T}^- \right)$, the chargino mass matrix is
\begin{equation*}
\Lc_{\text{Chargino Mass}}=-\frac{1}{2} (\psi^{\pm})^T \mathbf{M}_{\tilde{C}} \psi^{\pm}, \nn
\end{equation*}
where
\begin{equation*}
\mathbf{M}_{\tilde{C}}=
\begin{pmatrix}
\mathbf{0} & \mathbf{X}^T \\ \mathbf{X} & \mathbf{0}
\end{pmatrix}
 \nonumber, \\
\end{equation*}
 and
 \begin{equation}
 \mathbf{X}=
 \begin{pmatrix}
 M_{2}              &                         gvs_{\beta} &     -\sqrt{2}gv_{\chi} \\
    gvc_{\beta}    & \mu & 0 \\
-\sqrt{2}gv_{T}    &                 \sqrt{2}\lambda vs_{\beta} & \mu_{\Sigma}
 \end{pmatrix}.\label{eqn:charginoMixY1}
 \end{equation}
Finally, the doubly-charged fermion mass matrix is  
\begin{equation}
\Lc_{\text{Doubly Charged}} = -\frac{1}{2}\begin{pmatrix} \widetilde{\chi}^{++} & \widetilde{T}^{--} \end{pmatrix} \begin{pmatrix} 0 & -\mu_{\Sigma} \\ -\mu_{\Sigma} & 0 \end{pmatrix} \begin{pmatrix} \widetilde{\chi}^{++} \\ \widetilde{T}^{--} \end{pmatrix}.
\end{equation}

\bibliography{MyBib}

\end{document}